\documentclass{article}

\usepackage{float}
\usepackage{pdflscape}
\usepackage{longtable}

\usepackage{enumitem}
\usepackage{arxiv}
\usepackage{subcaption}
\usepackage{multirow}
\usepackage{amsmath}
\usepackage[utf8]{inputenc} 
\usepackage[T1]{fontenc}    
\usepackage{hyperref}       
\usepackage{url}            
\usepackage{booktabs}       
\usepackage{amsfonts}       
\usepackage{nicefrac}       
\usepackage{microtype}      
\usepackage{lipsum}
\usepackage{graphicx}
\graphicspath{ {./images/} }
\title{Pix2Geomodel: A Next-Generation Reservoir Geomodeling with Property-to-Property Translation}

\author{
 Abdulrahman Al-Fakih* \\
  College of Petroleum Engineering and Geosciences\\
  King Fahd University of Petroleum \& Minerals\\
  Dhahran 31261, Saudi Arabia \\
  \texttt{alfakihabdulrahman2030@gmail.com} \\
   \And
   Ardiansyah Koeshidayatullah \\
  College of Petroleum Engineering and Geosciences\\
  King Fahd University of Petroleum \& Minerals\\
  Dhahran 31261, Saudi Arabia \\
  \texttt{a.koeshidayatullah@kfupm.edu.sa} \\
  \And
    Nabil A. Saraih \\
  College of Petroleum Engineering and Geosciences\\
  King Fahd University of Petroleum \& Minerals\\
  Dhahran 31261, Saudi Arabia \\
  \texttt{nabil.saraih@kfupm.edu.sa} \\
  \And
 Tapan Mukerji \\
  Departments of Energy Science \& Engineering,\\ Earth \& Planetary Sciences, and Geophysics\\
  University, Stanford, CA, USA\\
  \texttt{mukerji@stanford.edu} \\
  \And
    Rayan Kanfar \\
  EXPEC Advanced Research Center\\
  Saudi Aramco\\Dhahran, Saudi Arabia \\
  \texttt{rayan.kanfar@aramco.com} \\
  \And
   Abdulmohsen Alali \\
  EXPEC Advanced Research Center\\
  Saudi Aramco\\Dhahran, Saudi Arabia \\
  \texttt{abdulmohsen.ali.1@aramco.com} \\
  \And
SanLinn I. Kaka \\
  College of Petroleum Engineering and Geosciences\\
  King Fahd University of Petroleum \& Minerals\\
  Dhahran 31261, Saudi Arabia \\
  \texttt{skaka@kfupm.edu.sa} \\
}

\begin{document}
\maketitle
\begin{abstract}
Accurate geological modeling is critical for reservoir characterization, yet traditional methods struggle with complex subsurface heterogeneity, and they have problems with conditioning to observed data. This study introduces Pix2Geomodel, a novel conditional generative adversarial network (cGAN) framework based on Pix2Pix, designed to predict reservoir properties (facies, porosity, permeability, water saturation) from the Groningen gas field’s Rotliegend reservoir. Utilizing a 7.6 million-cell dataset from the Nederlandse Aardolie Maatschappij, accessed via EPOS-NL, the methodology included data preprocessing, augmentation to generate 2,350 images per property, and training with a U-Net generator and PatchGAN discriminator over 19,000 steps. Evaluation metrics include pixel accuracy (PA), mean intersection over union (mIoU), frequency-weighted intersection over union (FWIoU), and visualizations assessed performance in masked property prediction and property-to-property translation tasks. Results demonstrated high accuracy for facies (PA 0.88, FWIoU 0.85) and water saturation (PA 0.96, FWIoU 0.95), with moderate success for porosity (PA 0.70, FWIoU 0.55) and permeability (PA 0.74, FWIoU 0.60), and robust translation performance (e.g., facies-to-Sw PA 0.98, FWIoU 0.97). The framework captured spatial variability and geological realism, as validated by variogram analysis, and calculated the training loss curves for the generator and discriminator for each property. Compared to traditional methods, Pix2Geomodel offers enhanced fidelity in direct property mapping. Limitations include challenges with microstructural variability and 2D constraints, suggesting future integration of multi-modal data and 3D modeling (Pix2Geomodel v2.0). This study advances the application of generative AI in geoscience, supporting improved reservoir management and open science initiatives.
\end{abstract}

\section*{Keywords}
Generative adversarial networks models; Time series generative adversarial networks models; Sequence generative adversarial networks models; Well log data imputation; Synthetic well log data generation.
\section*{Nomenclature}

\begin{tabbing}
\hspace{4cm} \= \hspace{3cm} \= \kill
Geomodelling \> \> Geological modeling \\
CCS \> \> Carbon capture and storage \\
ML \> \> Machine Learning \\
DL \> \> Deep Learning \\
GANs \> \> Generative Adversarial Networks \\
CGAN \> \> Conditional Generative Adversarial Networks \\
PixelRNN \> \> Pixel Recurrent Neural Network \\
AutoEncoder\> \> Autoencoding Neural Network \\
VAE\> \> Variational Autoencoder \\
CycleGAN \> \> Cycle-Consistent Generative Adversarial Network \\
StyleGAN \> \> Style-Based Generative Adversarial Network \\
iTiT \> \>Image-to-Image Translation  \\
Pix2Pix GAN \> \> Conditional Image-to-Image Translation Generative Adversarial Network \\
GANSim \> \> Conditional Facies Simulation using Generative Adversarial Networks \\
SPADE-GAN \> \> Spatially-Adaptive Denormalization Generative Adversarial Network \\
Stochastic Pix2Vid \> \> Spatiotemporal Image-to-Video Synthesis using Pix2Pix GAN Framework \\
SSIM \> \> Structural Similarity Index Measure (used for structural consistency tests) \\
Q-Q divergence \> \> Quantile-Quantile Divergence (statistical measure of distribution similarity) \\
GenAI \> \> Generative Artificial Intelligence \\
NAM\> \> Nederlandse Aardolie Maatschappij \\
EPOS-NL initiative \> \> European Plate Observing System – Netherlands initiative \\
Gslib format \> \> Geostatistical Software Library data format \\
I, J, K \> \> Grid cell indices for the X, Y, and Z directions in a 3D reservoir model \\
CSV \> \> Comma-Separated Values \\
EDA \> \> Exploratory Data Analysis \\
MAE \> \> Mean Absolute Error \\
L1 loss \> \> Mean Absolute Error (sum of absolute differences between predicted and actual values) \\
U-Net generator \> \> Encoder-decoder neural network with skip connections, used for image-to-image translation \\
PatchGAN \> \> Patch-based Generative Adversarial Network discriminator, classifies real/fake at patch level \\
BatchNorm \> \> Batch Normalization, stabilizes and accelerates neural network training \\
Transpose Conv \> \> Transposed Convolution (deconvolution), used for upsampling in neural networks \\
Tanh \> \> Hyperbolic tangent activation function \\
ReLU \> \> Rectified Linear Unit activation function \\
TensorBoard \> \> Visualization toolkit for monitoring deep learning training in TensorFlow \\
G \> \> Generator \\
D \> \> Discriminator \\
GPU \> \> Graphics Processing Unit \\
VRAM \> \> Video Random Access Memory \\
RAM \> \> Random Access Memory \\
Python \> \> High-level programming language for scientific computing and machine learning \\
TensorFlow \> \> Open-source deep learning framework developed by Google \\
PyTorch \> \> Open-source deep learning library developed by Facebook \\
pandas \> \> Python library for data manipulation and analysis \\
NumPy \> \> Python library for numerical computations \\
Matplotlib \> \> Python plotting library for data visualization \\\
mD \> \> Milli darcy (unit of permeability) \\
Sw \> \> Water saturation \\
PA \> \> Pixel Accuracy \\
mPA \> \> Mean Pixel Accuracy \\
mIoU \> \> Mean Intersection over Union \\
FWIoU \> \> Frequency-Weighted Intersection over Union \\

\end{tabbing}

\section{Introduction}
\begin{figure}[htbp] 
    \centering
    \includegraphics[width=1\textwidth]{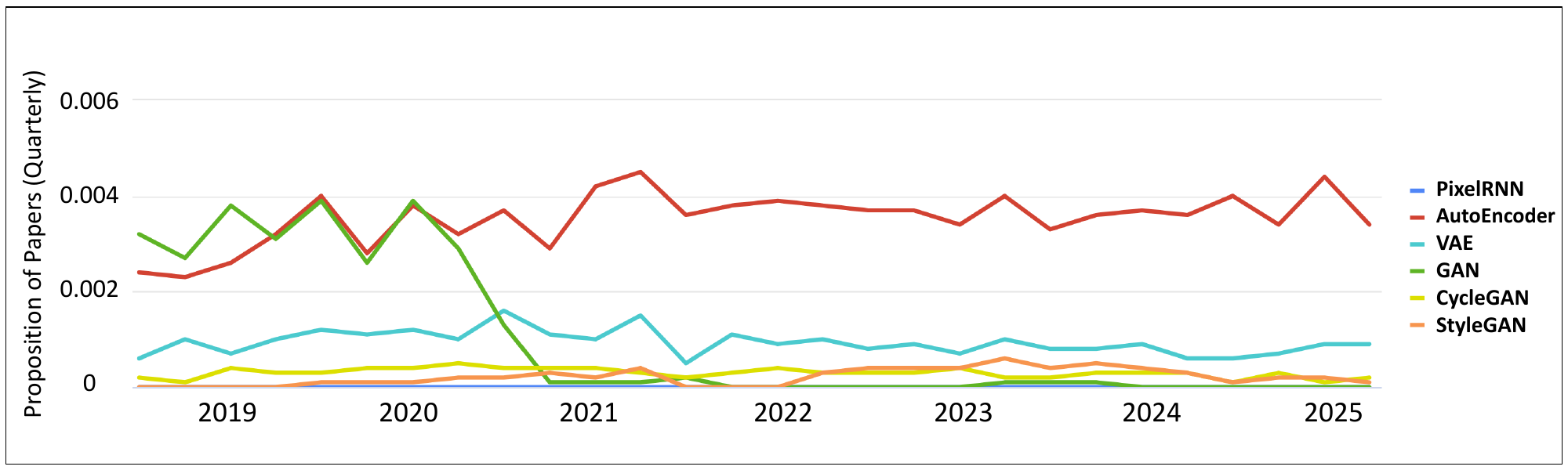}  
    \caption{Usage over time of generative models (2018-2025) [source: \url{https://www.paperswithcode.com/method/pixelrnn}].}
\label{fig:1}
\end{figure}

{Geological modeling (Geomodelling) is a fundamental process in subsurface characterization, essential for the exploration, development, and management of hydrocarbon reservoirs, groundwater resources, and carbon capture and storage (CCS) applications \cite{celia2009practical, pyrcz2014geostatistical, singh2014groundwater}. The primary goal of geomodeling is to construct accurate representations of subsurface formations, capturing their structural, stratigraphic pattern, and petrophysical properties \cite{pyrcz2014geostatistical, pandey2020geomodeling, garner2019closing, zhang2022three, cannon2018reservoir}. Facies models, which describe the spatial distribution of distinct rock types or sedimentary environments, are crucial as they directly influence fluid flow behavior and reservoir performance \cite{cavero2016importance, solanke2024techniques, zare2020reservoir}. Among the petrophysical properties, porosity and permeability are paramount, governing the storage and flow capacity of subsurface formations, thus making their accurate prediction vital for effective reservoir management and production optimization \cite{cavero2016importance, solanke2024techniques, okon2021artificial, zare2020reservoir}.

Traditional approaches to geomodeling, such as pixel-based or object-based modeling, rely on geostatistical methods to interpolate sparse well data or incorporate conceptual geological knowledge \cite{goovaerts1997geostatistics, cannon2018reservoir, obradors2023integrating}. While these methods have been widely adopted, they face significant limitations in capturing the complex, heterogeneous, and non-linear patterns inherent in subsurface systems \cite{zhang2023influence, mejia2024enhanced, khalili2023reservoir, catinat2023characterizing, yousefzadeh2023uncertainty}. For instance, pixel-based techniques often assume stationarity and struggle to model intricate spatial relationships \cite{kupfersberger1998deriving}, while object-based methods are less effective when conditioning to dense datasets \cite{strebelle2002conditional}. Moreover, predicting porosity and permeability from facies models is particularly challenging due to the complex interplay between depositional processes, diagenesis, and rock properties \cite{pyrcz2014geostatistical, wang2017identification}. These limitations underscore the need for advanced computational techniques capable of handling large datasets, modeling complex patterns, and improving prediction accuracy.

Recent advancements in machine learning (ML), particularly deep learning (DL), have provided new tools to address these challenges \cite{goodfellow2016deep}. Generative adversarial networks (GANs), introduced by Goodfellow et al. (2014), have emerged as a powerful framework for generating realistic data distributions and modeling complex systems. GANs consist of two neural networks, a generator that generates synthetic data from the real data and a discriminator that works similarly as a judge to evaluate its authenticity, trained simultaneously in a competitive setting \cite{goodfellow2014generative}. This framework has demonstrated remarkable success in domains such as image generation, natural language processing, and scientific modeling due to its ability to learn high-dimensional data distributions without explicit assumptions about their structure \cite{radford2015unsupervised, arjovsky2017wasserstein}.

A particularly promising development is image-to-image translation (iTiT), which transforms images from one domain to another while preserving their structural integrity \cite{isola2017image}. The Pix2Pix framework, a conditional generative adversarial network (cGAN) introduced by Isola et al. (2017), has set a benchmark for iTiT by enabling high-quality paired translations, such as converting sketches to realistic images or satellite imagery to street maps \cite{zhu2017unpaired, isola2017image}. Building on foundational work in texture synthesis and convolutional neural networks \cite{efros1999texture, efros2023image, gatys2016image}, Pix2Pix has inspired advanced variants (Fig.~\ref{fig:1}) like Pix2Pix-zero, differential image Pix2Pix, and multi-scale gradient U-Net, broadening its applications in fields such as medical imaging, urban planning, and autonomous vehicle training \cite{chen2009sketch2photo, shih2013data, gao2021survey, henry2021pix2pix, senapati2024image, wang2021multi, li2024mapping, li2023pix2pix}}.

In geoscience, GANs have shown great promise in geological analysis by enhancing seismic interpretation accuracy \cite{zhong2020predicting}, generating synthetic seismic datasets for improved fault detection model training \cite{choi2025synthetic}, and providing realistic velocity models to facilitate DL inversion workflows \cite{parasyris2023synthetic}. Building on these advancements, innovative GAN applications have expanded significantly, demonstrating their versatility in interpreting complex subsurface data. For instance, Weldemikael et al. (2024) effectively utilized(conditional image-to-image translation)Pix2Pix GAN frameworks to synthesize land gravity anomalies from satellite-based observations, greatly improving the spatial accuracy and reliability of geophysical interpretations \cite{weldemikael2024generating}. Baharuddin \& John (2025) introduced a novel application by translating microresistivity image logs into synthetic core images, accurately replicating vital geological features such as Dunham textures \cite{baharuddin2025using}. Moreover, Shahnas et al. (2025) leveraged GANs to process and invert geoelectrical field data, providing robust solutions to traditional inversion challenges in subsurface exploration \cite{shahnas2025application}.

In geological geomodelling, GANs have shown great promise as well by driving advances in stochastic facies modeling, porous‐media reconstruction, reservoir facies synthesis from training images and channelized conditioning, fully volumetric history matching, high‐fidelity uncertainty‐aware facies generation (including non‐stationary and 3D conditioning), surrogate‐based facies simulation, interpretable ensemble history matching in data‐sparse settings, spatiotemporal CO$_2$ plume synthesis, and the development of rigorous acceptance criteria for generated subsurface models. For a detailed breakdown of each method’s role, contributions, strengths, limitations, and emerging themes, see Table~\ref{Table1}.

Despite these advancements in GAN-based geological analysis, iTiT applications in geological modeling remain limited, with traditional algorithms for tasks like edge detection \cite{xie2015holistically} or semantic labeling \cite{eigen2015predicting} struggling to capture subsurface heterogeneity \cite{hertzmann2001image}. While stochastic and 3D adaptations of Pix2Pix have shown promise in history matching and model conditioning \cite{luy2021pix2pix, pan2021stochastic}, they do not directly address the prediction of paramount reservoir properties like porosity and permeability.

To bridge this gap, we introduce Pix2Geomodel, a novel conditional GAN framework inspired by Pix2Pix, specifically designed for geological datasets. Pix2Geomodel is the first application of the Pix2Pix framework in reservoir geomodeling, enabling the direct translation of facies maps into high-resolution predictions of porosity, permeability, and water saturation, and even the reverse, from porosity to facies. This framework aims to surpass the limitations of traditional methods by providing a scalable, data-driven approach to modeling subsurface heterogeneity and highlighting the transformative potential of generative AI in geoscience.

\begin{landscape}
\fontsize{8}{8.5}\selectfont
\begin{longtable}{|p{2.7cm}|p{2.2cm}|p{2.7cm}|p{3.6cm}|p{2.8cm}|p{2.9cm}|p{2.8cm}|}
\caption{
Overview of foundational and state-of-the-art GAN-based geomodeling studies (2017–2025), summarizing each work’s method/variant, primary application, conceptual contribution, key strengths and limitations, and emerging research themes.
} \\
\hline
\textbf{Ref.} & \textbf{Method / GAN Variant} & \textbf{Geomodeling Application} & \textbf{Conceptual Contribution} & \textbf{Strengths} & \textbf{Limitations} & \textbf{Potential Themes} \\
\hline

\cite{chan2017parametric}; 
\cite{mosser2017reconstruction};
\cite{mosser2018conditioning};
\cite{mosser2019deepflow} 
& GAN 
& \begin{itemize}[leftmargin=*, itemsep=0pt, labelsep=3pt, parsep=0pt, topsep=0pt, partopsep=0pt]
    \item Stochastic facies modeling
    \item Reconstruction of porous media from limited data
  \end{itemize}
& Demonstrated the feasibility of GANs to generate realistic subsurface heterogeneity in 2D facies distributions and 3D pore-/reservoir-scale structures from sparse observations.
& \begin{itemize}[leftmargin=*, itemsep=0pt, labelsep=3pt, parsep=0pt, topsep=0pt, partopsep=0pt]
    \item First proof-of-concept for GANs in facies simulation
    \item Generated realistic 3D reconstructions under data scarcity
  \end{itemize}
& \begin{itemize}[leftmargin=*, itemsep=0pt, labelsep=3pt, parsep=0pt, topsep=0pt, partopsep=0pt]
    \item Limited to simple facies scenarios / narrow geological settings
    \item High computational cost for high-resolution volumes
  \end{itemize}
& \begin{itemize}[leftmargin=*, itemsep=0pt, labelsep=3pt, parsep=0pt, topsep=0pt, partopsep=0pt]
    \item Early generative subsurface modeling
    \item Data-driven volumetric reconstruction
  \end{itemize}
\\ \hline

\cite{liu2021improved}; \cite{pan2021stochastic}
& GAN; Stochastic Pix2Pix
& \begin{itemize}[leftmargin=*, itemsep=0pt, labelsep=3pt, parsep=0pt, topsep=0pt, partopsep=0pt]
    \item Reservoir facies modeling from geological training images
    \item Conditioning channelized reservoir models
  \end{itemize}
& Bridged traditional geostatistics and DL by synthesizing facies from training images and mapping geophysical + well-log inputs to channelized facies via a conditional Pix2Px.
& \begin{itemize}[leftmargin=*, itemsep=0pt, labelsep=3pt, parsep=0pt, topsep=0pt, partopsep=0pt]
    \item Improved geological realism vs. conventional geostatistics
    \item Integrated multiple data sources for accurate heterogeneity
  \end{itemize}
& \begin{itemize}[leftmargin=*, itemsep=0pt, labelsep=3pt, parsep=0pt, topsep=0pt, partopsep=0pt]
    \item Dependent on training-image quality
    \item Risk of overfitting
    \item Increased training complexity with multi-source inputs / limited 3D demonstration
  \end{itemize}
& \begin{itemize}[leftmargin=*, itemsep=0pt, labelsep=3pt, parsep=0pt, topsep=0pt, partopsep=0pt]
    \item Conditioning on geological priors
    \item Multi-source data fusion
  \end{itemize}
\\ \hline

\cite{luy2021pix2pix}
& 3D Pix2Pix GAN
& History matching of reservoir parameters
& Extended Pix2Pix into 3D, enabling GAN-based history matching by aligning dynamic performance data with a fully volumetric facies model, thereby integrating temporal observations directly into geomodeling.
& \begin{itemize}[leftmargin=*, itemsep=0pt, labelsep=3pt, parsep=0pt, topsep=0pt, partopsep=0pt]
    \item Incorporated temporal (history) data into geomodels
    \item Generated full-volume 3D facies consistent with production history
  \end{itemize}
& \begin{itemize}[leftmargin=*, itemsep=0pt, labelsep=3pt, parsep=0pt, topsep=0pt, partopsep=0pt]
    \item Very high computational cost for 3D training
    \item Requires substantial dynamic observation data
  \end{itemize}
& \begin{itemize}[leftmargin=*, itemsep=0pt, labelsep=3pt, parsep=0pt, topsep=0pt, partopsep=0pt]
    \item Dynamic data assimilation (history matching)
    \item 3D volumetric GANs
  \end{itemize}
\\ \hline

\cite{song2021gansim}
& GANSim
& High-fidelity facies modeling with uncertainty quantification
& Introduced conditional facies simulation GAN (GANSim), a GAN framework tailored to geological facies that explicitly quantifies uncertainty in generated distributions, improving on earlier GANs by providing uncertainty measures alongside realism.
& \begin{itemize}[leftmargin=*, itemsep=0pt, labelsep=3pt, parsep=0pt, topsep=0pt, partopsep=0pt]
    \item Enhanced facies realism over prior methods
    \item Provided explicit uncertainty metrics
  \end{itemize}
& \begin{itemize}[leftmargin=*, itemsep=0pt, labelsep=3pt, parsep=0pt, topsep=0pt, partopsep=0pt]
    \item Computationally demanding for large 3D grids
    \item Relies on representative training sets
  \end{itemize}
& \begin{itemize}[leftmargin=*, itemsep=0pt, labelsep=3pt, parsep=0pt, topsep=0pt, partopsep=0pt]
    \item Uncertainty quantification in GANs
    \item Geological conditioning
  \end{itemize}
\\ \hline

\cite{zhang2019generating};\cite{chen2022modeling};\cite{song2022gansim3d}
& GAN with self-attention; GANSim-3D
& \begin{itemize}[leftmargin=*, itemsep=0pt, labelsep=3pt, parsep=0pt, topsep=0pt, partopsep=0pt]
    \item Modeling non-stationary facies distributions
    \item 3D facies modeling with conditioning
  \end{itemize}
& Incorporated self-attention to capture global spatial dependencies in complex depositional and extended GANSim into full 3D volumes (GANSim-3D), preserving vertical/lateral heterogeneity.
& \begin{itemize}[leftmargin=*, itemsep=0pt, labelsep=3pt, parsep=0pt, topsep=0pt, partopsep=0pt]
    \item Captured long-range correlations \& non-stationarity
    \item Produced realistic 3D facies honoring well data
  \end{itemize}
& \begin{itemize}[leftmargin=*, itemsep=0pt, labelsep=3pt, parsep=0pt, topsep=0pt, partopsep=0pt]
    \item Increased model complexity and training instability
    \item Very high computational cost \& memory footprint for full 3D
  \end{itemize}
& \begin{itemize}[leftmargin=*, itemsep=0pt, labelsep=3pt, parsep=0pt, topsep=0pt, partopsep=0pt]
    \item Advanced architectures (attention)
    \item Volumetric heterogeneity
  \end{itemize}
\\ \hline

\cite{zhang2023geologically}; \cite{song2023gansim}
& GANSim (enhanced); GANSim-surrogate
& \begin{itemize}[leftmargin=*, itemsep=0pt, labelsep=3pt, parsep=0pt, topsep=0pt, partopsep=0pt]
    \item Facies distribution modeling with improved realism \& uncertainty
    \item Surrogate modeling for facies simulation
  \end{itemize}
& Enhanced GANSim to deliver higher-fidelity facies and more robust uncertainty metrics, and developed a surrogate network (GANSim-surrogate) to accelerate facies model generation for near real-time workflows.
& \begin{itemize}[leftmargin=*, itemsep=0pt, labelsep=3pt, parsep=0pt, topsep=0pt, partopsep=0pt]
    \item Superior geological realism and rigorous uncertainty quantification
    \item Significantly reduced computational time for generation
  \end{itemize}
& \begin{itemize}[leftmargin=*, itemsep=0pt, labelsep=3pt, parsep=0pt, topsep=0pt, partopsep=0pt]
    \item Further increased training complexity; may require larger datasets for reliable uncertainties
    \item Surrogate may oversmooth fine-scale features, introducing approximation errors
  \end{itemize}
& \begin{itemize}[leftmargin=*, itemsep=0pt, labelsep=3pt, parsep=0pt, topsep=0pt, partopsep=0pt]
    \item Iterative framework improvements
    \item Surrogate modeling for speed
  \end{itemize}
\\ \hline

\cite{fossum2024ensemble}; \cite{ranazzi2024improving}
& SPADE-GAN: Adaptive GAN with discriminator regularization
& \begin{itemize}[leftmargin=*, itemsep=0pt, labelsep=3pt, parsep=0pt, topsep=0pt, partopsep=0pt]
    \item Ensemble history-matching workflows
    \item Reservoir model generation under limited data conditions
  \end{itemize}
& Introduced interpretable spatially-adaptive denormalization (SPADE)-GANs for ensemble adjustments to geomodels based on observed data (Fossum et al.) and proposed adaptive training strategies (zero-centered discriminator regularization + augmentation) for sparse-data robustness (Ranazzi et al.).
& \begin{itemize}[leftmargin=*, itemsep=0pt, labelsep=3pt, parsep=0pt, topsep=0pt, partopsep=0pt]
    \item Combined interpretability with generative power
    \item Enabled robust training under very limited datasets
  \end{itemize}
& \begin{itemize}[leftmargin=*, itemsep=0pt, labelsep=3pt, parsep=0pt, topsep=0pt, partopsep=0pt]
    \item Interpretability vs. generative flexibility trade-offs; complex implementation
    \item Potential over-regularization, suppressing fine details
    \item Requires domain-specific tuning
  \end{itemize}
& \begin{itemize}[leftmargin=*, itemsep=0pt, labelsep=3pt, parsep=0pt, topsep=0pt, partopsep=0pt]
    \item Interpretable GANs
    \item Training stability under data scarcity
  \end{itemize}
\\ \hline

\cite{morales2024stochastic}
& Stochastic Pix2Vid
& Spatiotemporal geomodel synthesis for CO$_2$ storage
& Extended Pix2Pix into a spatiotemporal “video” framework (Stochastic Pix2Vid), modeling CO$_2$ plume evolution over time and enabling predictive scenarios for subsurface processes.
& \begin{itemize}[leftmargin=*, itemsep=0pt, labelsep=3pt, parsep=0pt, topsep=0pt, partopsep=0pt]
    \item Captured the temporal evolution of geological processes
    \item Enabled predictive spatiotemporal modeling
  \end{itemize}
& \begin{itemize}[leftmargin=*, itemsep=0pt, labelsep=3pt, parsep=0pt, topsep=0pt, partopsep=0pt]
    \item High dimensionality leads to greater training times
    \item Demonstrated primarily on CO$_2$storage, limiting broader generalization
  \end{itemize}
& \begin{itemize}[leftmargin=*, itemsep=0pt, labelsep=3pt, parsep=0pt, topsep=0pt, partopsep=0pt]
    \item Spatiotemporal forecasting
    \item Predictive geological evolution
  \end{itemize}
\\ \hline

\cite{merzoug2024conditional}
& Quantitative evaluation framework
& Acceptance criteria \& quantitative checks for conditional GAN geomodels
& Provided a standardized, domain-focused suite of quantitative checks, histogram/quantile-quantile divergence (Q-Q) divergence, Earth Mover’s Distance, variogram analyses, structural consistency tests (SSIM stability), conditioning metrics, and flow simulation recovery factors, to rigorously evaluate generative artificial intelligence (GenAI)-based subsurface models.
& \begin{itemize}[leftmargin=*, itemsep=0pt, labelsep=3pt, parsep=0pt, topsep=0pt, partopsep=0pt]
    \item Comprehensive evaluation protocol enhances model reliability and uncertainty awareness
  \end{itemize}
& \begin{itemize}[leftmargin=*, itemsep=0pt, labelsep=3pt, parsep=0pt, topsep=0pt, partopsep=0pt]
    \item Adds computational overhead to workflows
    \item Potential need for adaptation across varied geological contexts
  \end{itemize}
& \begin{itemize}[leftmargin=*, itemsep=0pt, labelsep=3pt, parsep=0pt, topsep=0pt, partopsep=0pt]
    \item Evaluation standards; quality control
    \item Uncertainty-aware genAI models
  \end{itemize}
\\ \hline

\cite{merzoug2025checking}
& Conditional GAN Evaluation Workflow
& Quantitative assessment of GAN-generated subsurface models
& Proposed minimum acceptance criteria workflow for cGAN-based subsurface modeling, focusing on (1) data distribution, (2) spatial continuity, and (3) local data conditioning; revealed that local conditioning introduces artifacts and requires more iterations as data points increase.
& Robust, quantitative framework for evaluating GAN outputs; highlights limitations in local conditioning
& \begin{itemize}[leftmargin=*, itemsep=0pt, labelsep=3pt, parsep=0pt, topsep=0pt, partopsep=0pt]
    \item Conditioning artifacts near data locations
    \item Increased error as conditioning data increases
    \item Additional evaluation overhead
  \end{itemize}
& Quantitative evaluation: conditioning challenges in cGAN workflows
\\ \hline
\end{longtable}
\label{Table1}
\end{landscape}

\section {Methodology}
\begin{figure}[htbp] 
    \centering
    \includegraphics[width=1\textwidth]{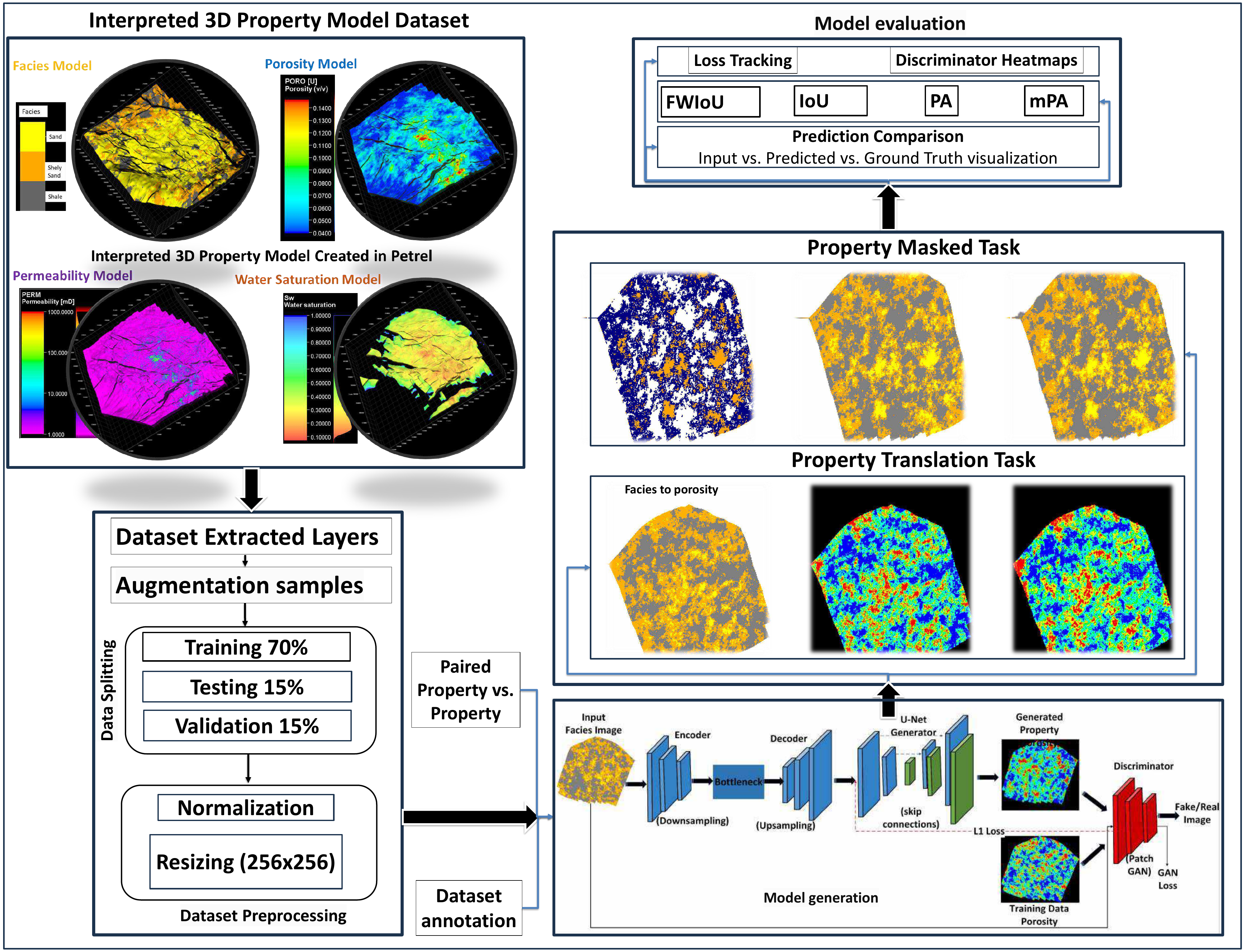}  
    \caption{High-level architecture of the proposed Pix2Geomodel framework for subsurface reservoir characterization, provides a high-level overview of the methodology, visually capturing the flow from dataset preparation to final evaluation.}
    \label{fig:2}
\end{figure}

The Pix2Geomodel framework leverages a cGAN based on the Pix2Pix architecture~\cite{isola2017image} to predict reservoir properties (facies, porosity, permeability, water saturation) for subsurface reservoir characterization. The methodology is organized into five key stages: dataset extraction and preprocessing, data augmentation, dataset structuring, model architecture and training, and evaluation. These stages ensure robust data preparation, model optimization, and performance assessment (Fig.~\ref{fig:2}).

\subsection{Dataset extraction and processing}

Reservoir property models were extracted from a three-dimensional geological model of the Groningen gas field’s Rotliegend reservoir, developed by the Nederlandse Aardolie Maatschappij (NAM) and accessed via the European Plate Observing System – Netherlands (EPOS-NL) initiative~\cite{matenco2023integrating}. Built in Petrel software, the model integrates 2D/3D seismic data, well logs, core samples, and production data within a grid framework of 523~$\times$~630~$\times$~235 cells (I, J, K). Each grid cell measures approximately 94.8 meters in the X-direction, 103.3 meters in the Y-direction, and 2.45 meters in the Z-direction. The model covers a subsurface depth range from a minimum of 2,543.6 meters to a maximum elevation depth of 4,543.42 meters. Property volumes (facies, porosity, permeability, water saturation) were exported in Gslib format using the ``Include cell index'' option to preserve I, J, K indices, then converted to comma-separated values (CSV) to facilitate easier manipulation and processing. Then, unnecessary headers were removed from Gslib files to streamline data handling. Inverse K-indices were corrected using the transformation $K_\text{corrected} = \max(K) - K_\text{original} + 1$ to properly align the depth axis from top to bottom.

The dataset is standardized by normalizing pixel intensity values, resizing images to a 256$\times$256 resolution, and performing additional augmentations to improve robustness during model training. This constitutes a supervised learning setup in the form of paired image-to-image translation. Each training sample consists of a source image (e.g., facies) and a corresponding target image (e.g., porosity). Unlike traditional supervised learning with scalar or tabular outputs, supervision here occurs at the pixel level, where the model learns to map spatial patterns from one property domain to another. The network is trained to minimize both adversarial and reconstruction losses against the ground truth, enabling Pix2Geomodel to capture complex spatial dependencies and produce geologically consistent outputs.

\begin{figure}[htbp] 
    \centering
    \includegraphics[width=1\textwidth]{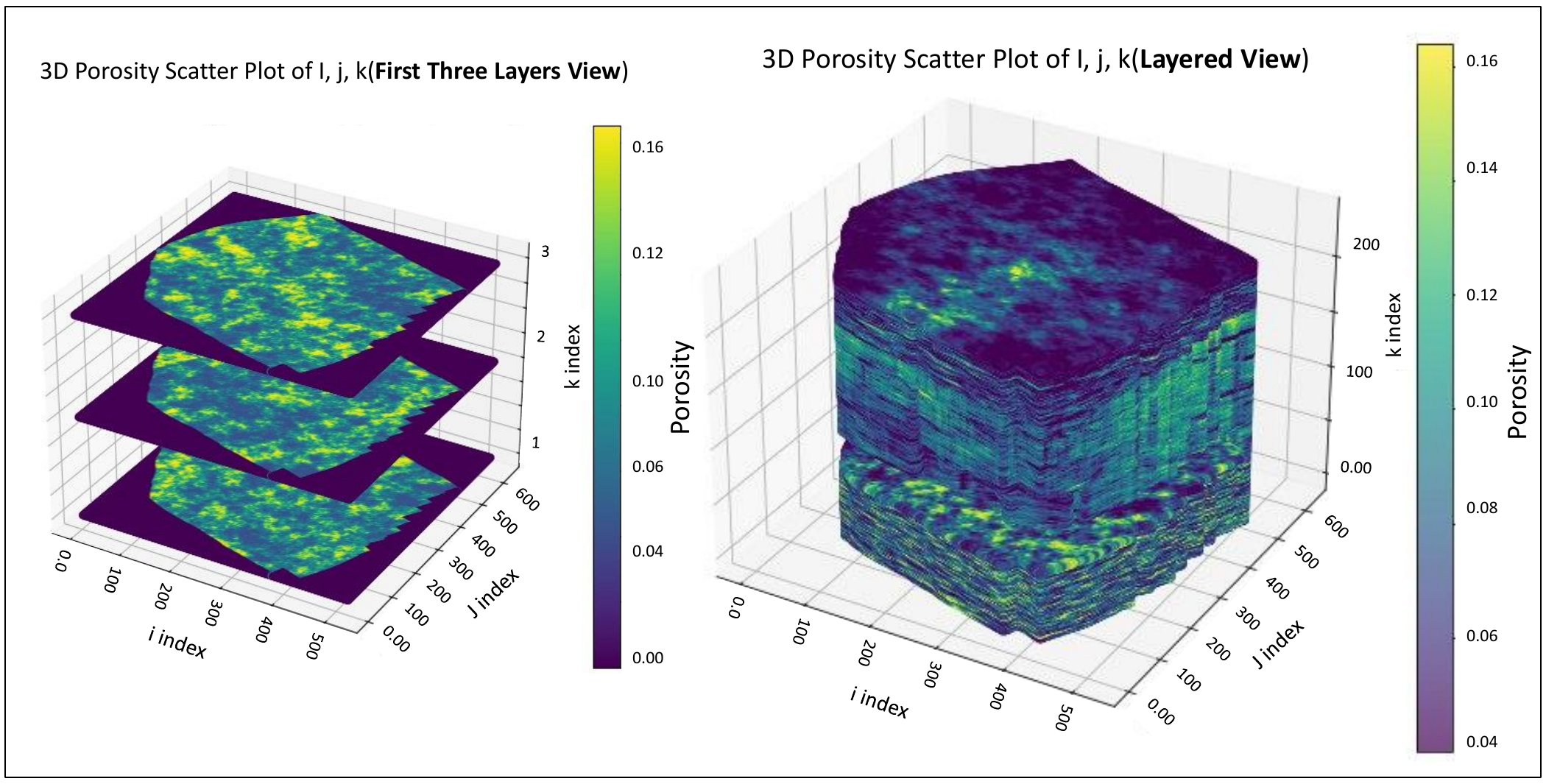}  
    \caption{Visualization of porosity distribution in a geological reservoir shown in two forms: the left-hand side illustrates a clear three-layered representation to highlight patterns within discrete geological layers; the right-hand side provides a full layered view emphasizing geological heterogeneity and structural continuity throughout the reservoir. Each layer is color-coded to reflect variations in porosity.}
\label{fig:3}   
\end{figure}

Pixel intensities were standardized to a [$-$1, 1] range to facilitate model training. High-resolution images (256~$\times$~256 pixels, 900 DPI) were generated by replicating Petrel’s visualization style through Python scripts that applied color-matching, contrast enhancement, and sharpness adjustments for visual fidelity. Exploratory data analysis (EDA) was conducted to verify data consistency, by visualizing 3D property distributions in two representations: the left-hand visualization depicts a three-layered example to explicitly show spatial patterns within specific geological layers, while the right-hand visualization shows a comprehensive layered view to emphasize geological heterogeneity and continuity across the entire reservoir (Fig.~\ref{fig:3}).

\subsection{Data enhancement techniques and configuration}

\begin{figure}[htbp] 
    \centering
    \includegraphics[width=1\textwidth]{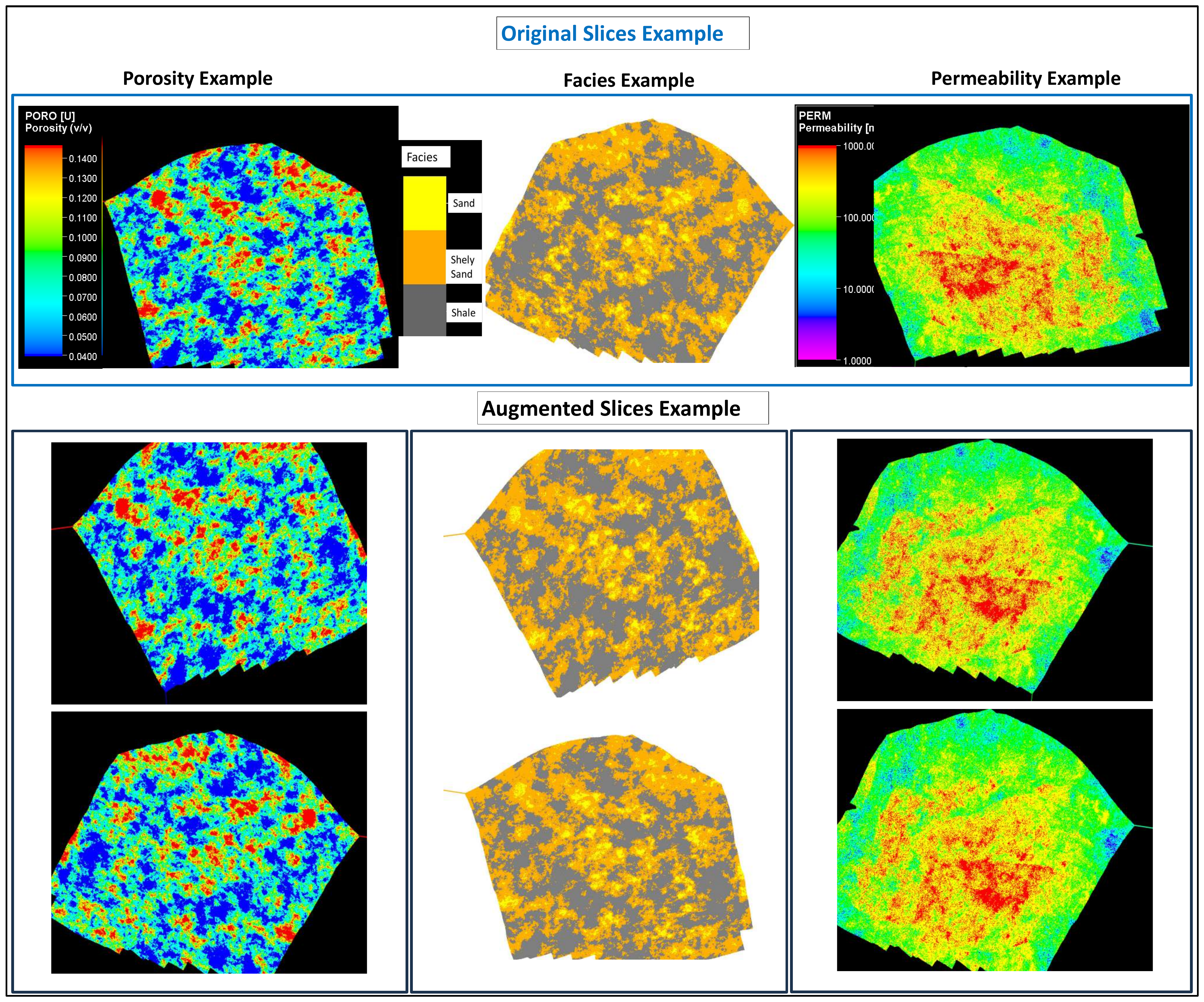}  
    \caption{Augmented property images generated for porosity using random transformations such as rotations, flipping, zooming, and translations. The examples showcase the diversity introduced through augmentation, improving model robustness while preserving geological consistency.}
\label{fig:4}   
\end{figure}

To enhance model generalization and mitigate overfitting, a robust data augmentation (Fig.~\ref{fig:4}) strategy was applied, generating 10 augmented versions of each of the 235 original layers per property, resulting in a total of 2,350 images per property. The augmentation process employed constrained transformations to maintain geological coherence, including rotations within $\pm$10$^\circ$, zooming ranging from 0.9$\times$ to 1.1$\times$, translations limited to $\pm$5\% of image dimensions, horizontal and vertical flipping, and random cropping from 286~$\times$~286 to 256~$\times$~256 pixels. Each augmented image underwent visual inspection to ensure the preservation of structural continuity and realistic spatial relationships, thereby upholding the geological integrity of the dataset (Fig.~\ref{fig:4}).

\begin{figure}[htbp]
    \centering
    \begin{subfigure}[t]{0.49\textwidth}
        \centering
        \includegraphics[width=\textwidth]{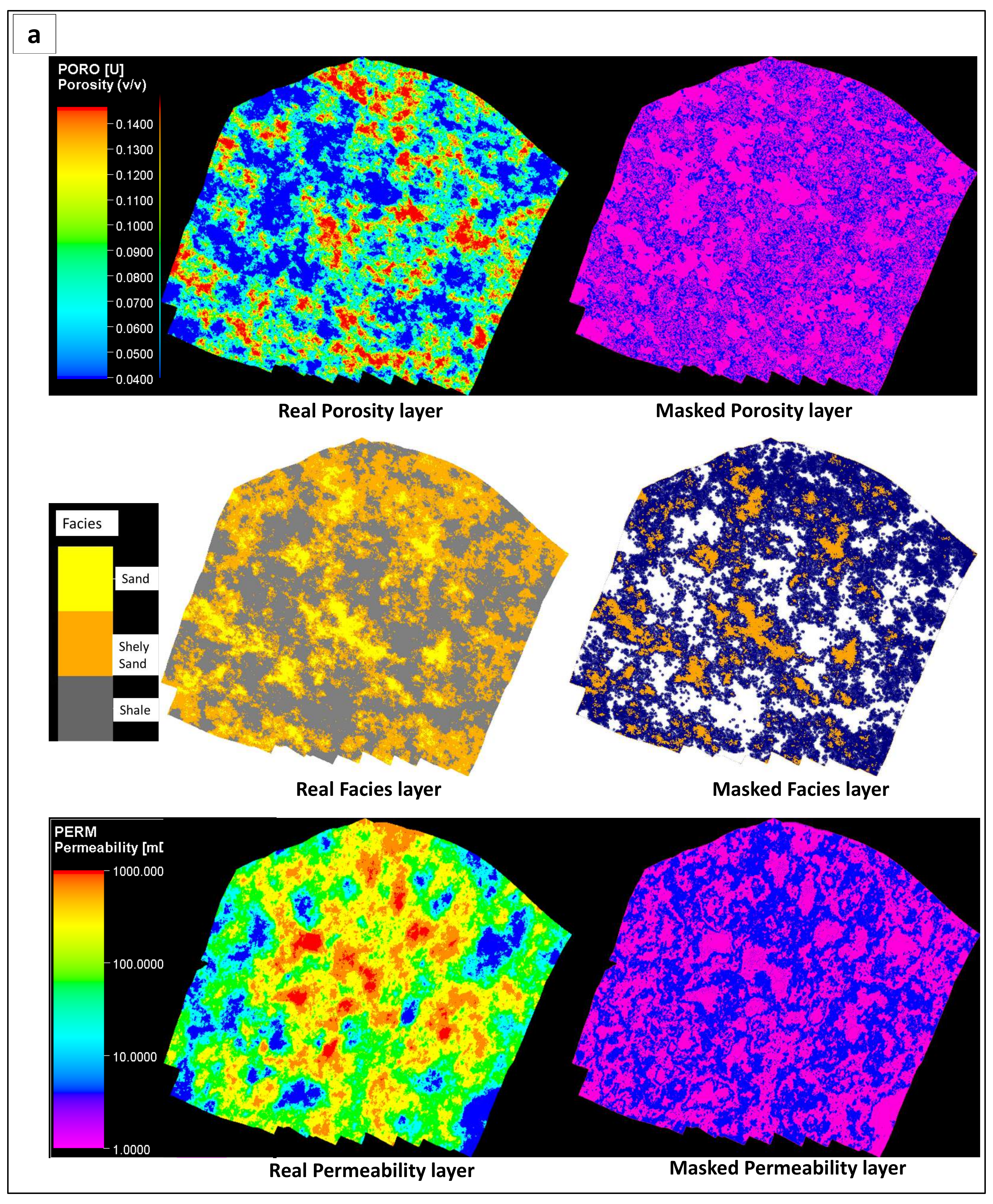}
      
        \label{fig:5a}
    \end{subfigure}
    \hfill
    \begin{subfigure}[t]{0.49\textwidth}
        \centering
        \includegraphics[width=\textwidth]{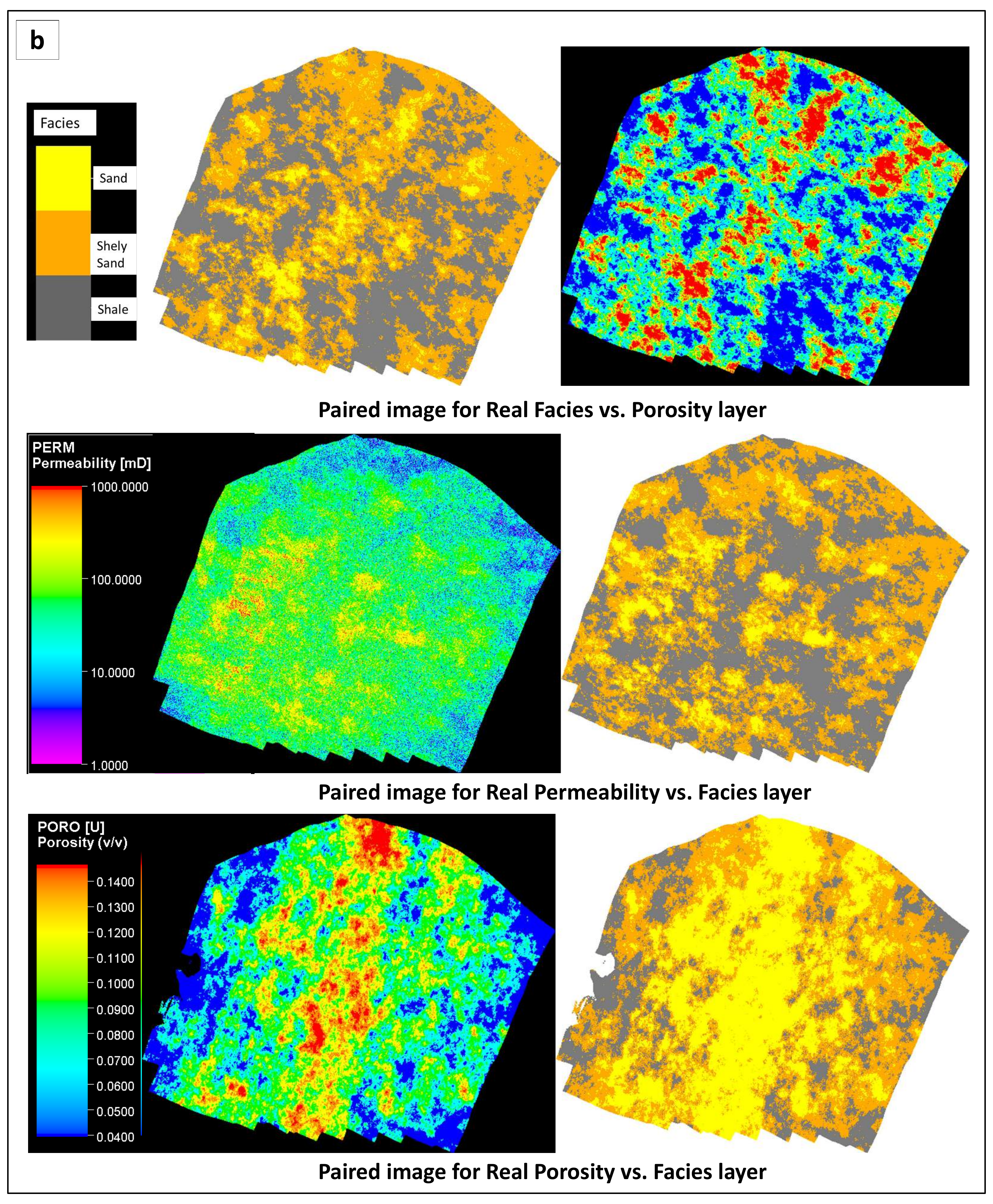}
      
        \label{fig:5b}
    \end{subfigure}
    \caption{Paired datasets for Pix2Geomodel tasks: (a) Masked property prediction with color-coded property masks paired with corresponding property images. Masking simulates incomplete geological data scenarios, testing Pix2Geomodel's ability to reconstruct missing or obscured geological information. (b) Facies-to-property translation where facies images are paired with property maps such as porosity, permeability, or Sw.}
    \label{fig:5}
\end{figure}

The dataset was structured to support two primary tasks: masked property prediction and property-to-property translation, with specific configurations to ensure robust training and evaluation (Fig.~\ref{fig:5}). For masked property prediction, an automated annotation process generated color-coded masks to simulate incomplete data scenarios, mapping pixel intensities to discrete classes and pairing these masks with corresponding property images such as porosity or permeability to form the input-output datasets required for training the Pix2Pix-based Pix2Geomodel framework (Fig.~\ref{fig:5}a). For property-to-property translation, input properties like facies were paired with target properties like porosity for scenarios including facies-to-porosity, porosity-to-facies, facies-to-permeability, and facies-to-water saturation, with paired datasets saved as PNG files after visual validation (Fig.~\ref{fig:5}b). The dataset was split into 70\% training (1,809 images), 15\% validation (389 images), and 15\% testing (390 images) per property, ensuring balanced representation across geological features.

\subsection{Pix2Geomodel architecture}

\begin{figure}[htbp] 
    \centering
    \includegraphics[width=1\textwidth]{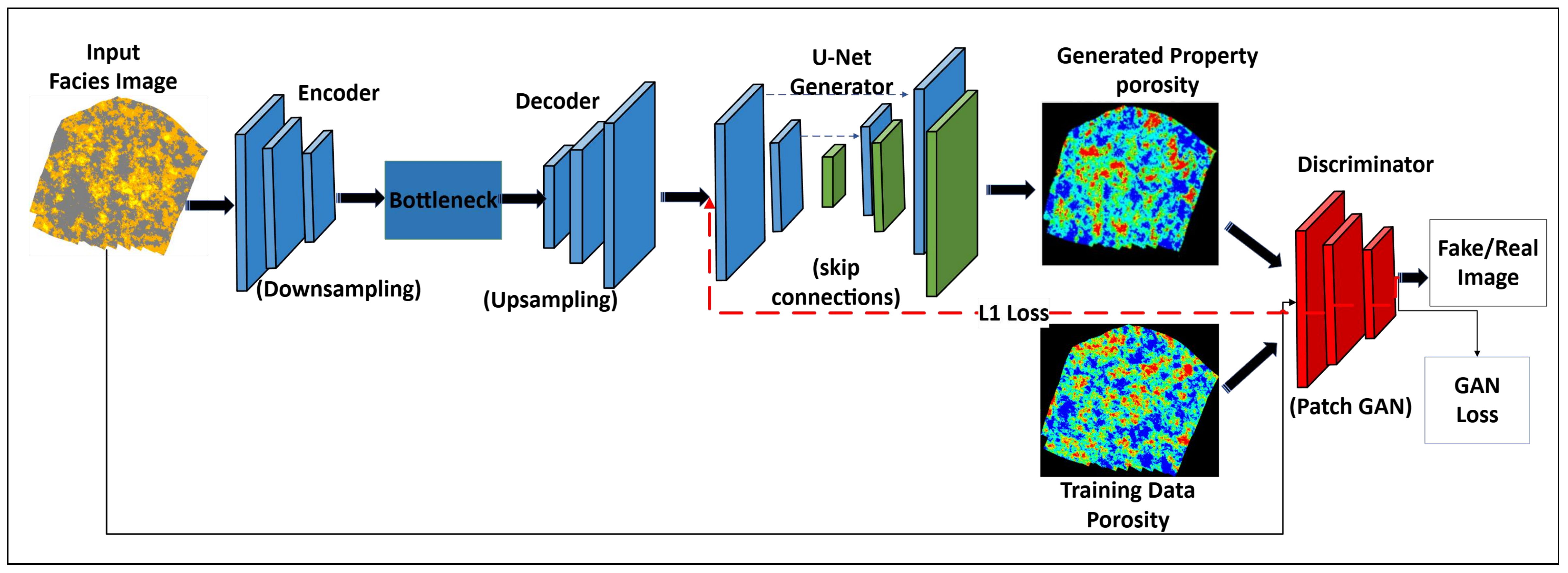}  
    \caption{ Architecture of the Pix2Geomodel framework, showcasing the overall architecture, including the interaction between the generator and discriminator.}
\label{fig:6}   
\end{figure}

\begin{figure}[htbp] 
    \centering
    \includegraphics[width=1.1\textwidth]{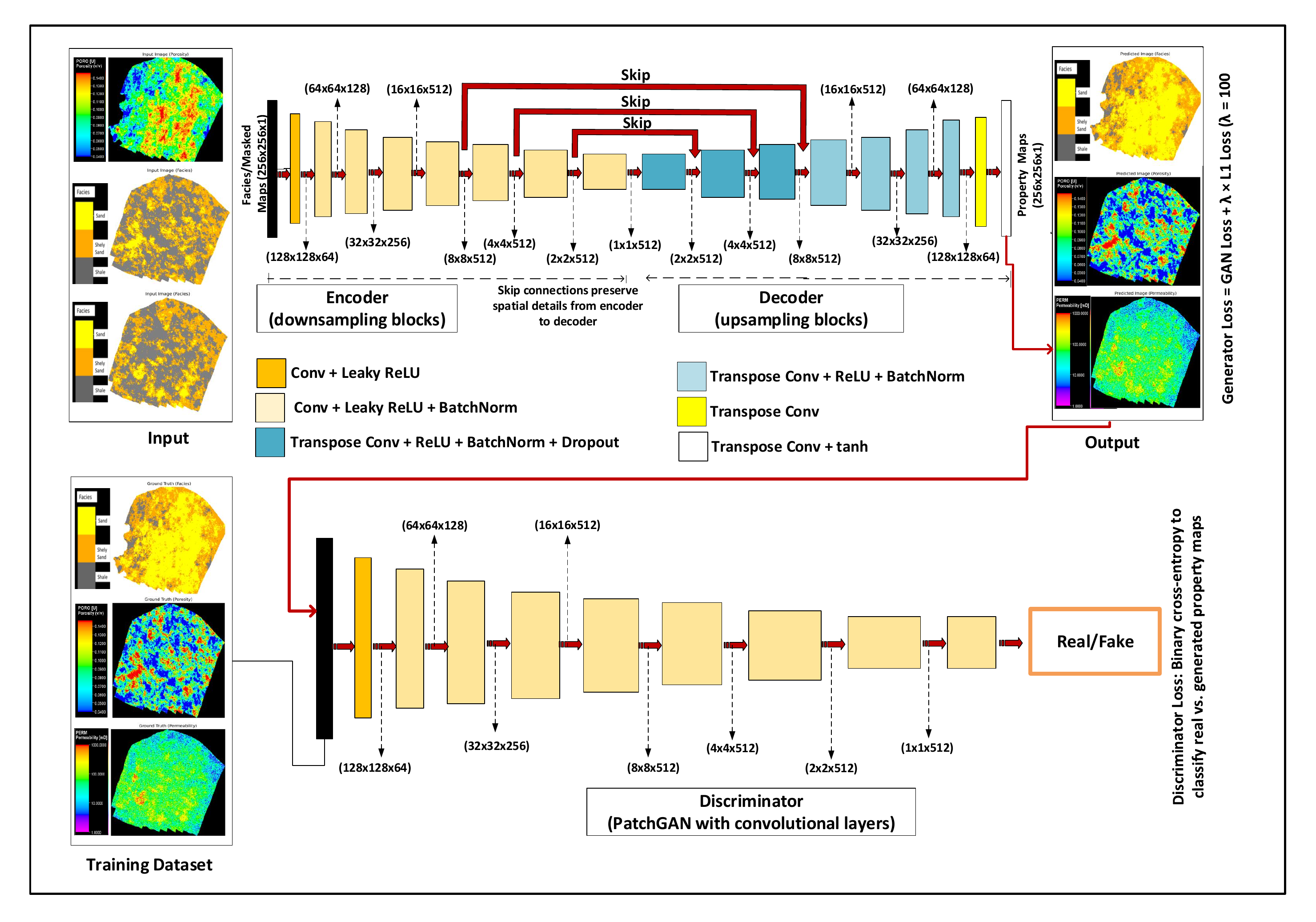}  
    \caption{ Schematic of the Pix2Geomodel for property-to-property translation. Top: U-Net generator with encoder–decoder blocks and skip connections. Bottom: PatchGAN discriminator composed of convolutional layers. The left shows the input property and training samples; the right displays the generated property maps.}
\label{fig:7}   
\end{figure}

The Pix2Geomodel framework employs a U-Net generator and PatchGAN discriminator, adapted from the Pix2Pix cGAN that was introduced by Isola et al.~\cite{isola2017image}, to generate high-fidelity property maps (Fig.~\ref{fig:6}, Fig.~\ref{fig:7}). The U-Net generator features an encoder that downsamples input images (256~$\times$~256~$\times$~1) through eight convolutional layers (128$\times$128~$\rightarrow$~64$\times$64~$\rightarrow$~32$\times$32~$\rightarrow$~16$\times$16~$\rightarrow$~8$\times$8~$\rightarrow$~4$\times$4~$\rightarrow$~2$\times$2~$\rightarrow$~1$\times$1) with Leaky ReLU activation and batch normalization, reducing dimensions to 1~$\times$~1, and a decoder that upsamples using transposed convolutions with ReLU activation, batch normalization, and applies 50\% dropout in the first three layers to regularize the network and prevent overfitting, followed by final \texttt{tanh} activation to restore outputs to 256~$\times$~256. Skip connections between encoder and decoder layers preserve spatial details lost during downsampling, enabling the generation of geologically accurate property maps. The PatchGAN discriminator evaluates 70~$\times$~70 patches of input-output pairs, using convolutional layers with Leaky ReLU and binary cross-entropy loss, reducing dimensions to 1~$\times$~1~$\times$~512 for classification, with its classification performance optimized through binary cross-entropy loss, ensuring the preservation of high-frequency details and textures key for geological accuracy.

\subsection{Training and optimization}

\begin{figure}[htbp] 
    \centering
    \includegraphics[width=1\textwidth]{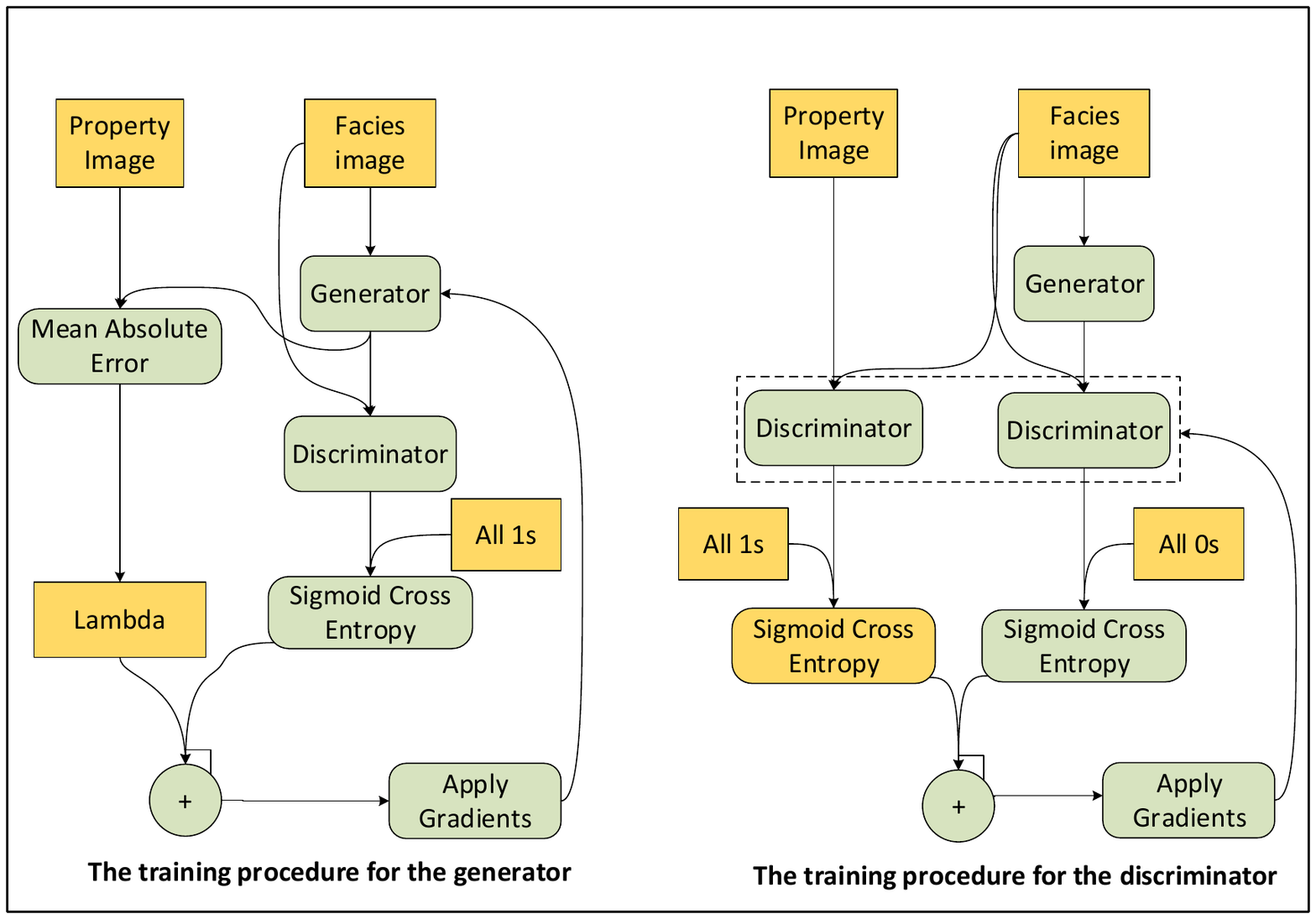}  
    \caption{ Training procedure of the Pix2Geomodel framework (adopted from Isola et al., 2017).}
\label{fig:8}   
\end{figure}

Training was conducted over 19,000 steps using a three-step process of forward pass, loss calculation, and backpropagation, optimizing the model for accurate and realistic property predictions (Fig.~\ref{fig:8}; Table~\ref{tab:2}). In the forward pass, the generator transformed input images, such as masked inputs or facies, into predicted property maps, which the discriminator evaluated against the ground truth/training property map to determine their authenticity. Loss calculation combined adversarial loss and L1 reconstruction loss. Adversarial loss defined as
\[
L_\mathrm{GAN} (G, D) = \mathbb{E}_{x, y} [\log D (x, y)] + \mathbb{E}_{x, z} [\log (1 - D(x, G(x, z)))]
\tag{1}
\]
where $G$ is the generator responsible for producing outputs $G(x, z)$ conditioned on the input image $x$ and noise $z$, and $D$ is the discriminator tasked with distinguishing real data pairs $(x, y)$ from generated pairs $(x, G(x, z))$. This adversarial loss encourages the generator to create outputs that are indistinguishable from real data, while the discriminator learns to classify pairs as real or fake effectively. On the other hand, L1 reconstruction loss defined as
\[
L_1(G) = \mathbb{E}_{x, y, z} [\|y - G(x, z)\|_1]
\tag{2}
\]
The L1 loss ensures that the generated image $G(x, z)$ is structurally similar to the ground truth image $y$, penalizing deviations and reducing artifacts such as blurring~\cite{ioffe2015batch}. The Pix2Pix framework integrates adversarial and reconstruction losses into a unified optimization objective, expressed as
\[
G^* = \arg\min_G \max_D L_\mathrm{GAN} (G, D) + \lambda L_1 (G)
\tag{3}
\]
with $\lambda = 100$ balancing realism and accuracy, as per Isola et al.~\cite{isola2017image}. Testing values like 10, 50, 200, and 1000 showed lower values caused underfitting and higher values over-smoothed outputs, confirming $\lambda = 100$ as optimal for visual fidelity and structural coherence. The discriminator loss measures the discriminator's accuracy in distinguishing real images from generated ones, employing sigmoid cross-entropy to assess separate loss components for real and fake inputs, thereby enhancing its capability to differentiate between authentic and synthetic property maps. Backpropagation computed gradients, optimized via the Adam optimizer (learning rate: 2e-4, beta1: 0.5), with a batch size of 1 and buffer size of 400, logging loss metrics every 10 steps and visualizing outputs every 1,000 steps via TensorBoard (Table~\ref{tab:2}).

The Pix2Geomodel framework optimizes performance using two key loss functions: the generator loss, which combines adversarial loss for realism and L1 loss for structural accuracy, expressed as 
\[
L_\mathrm{Generator} = L_\mathrm{GAN} + \lambda L_1
\tag{4}
\]
with $\lambda = 100$ to prioritize reconstruction, and the discriminator loss, which employs binary cross-entropy to assess the discriminator’s ability to distinguish real from generated property maps. Utilizing an encoder-decoder architecture with skip connections and adversarial training, Pix2Geomodel generates realistic geological property maps, with the target property (e.g., porosity, permeability) determined by selecting specific input-output image pairs during training, allowing each model to specialize in translating input facies maps into a designated property. This design, detailed in the architecture overview (Figs.~\ref{fig:6},~\ref{fig:7}), ensures that each model is independently trained on dedicated paired datasets to capture the unique characteristics of properties like porosity or water saturation, enhancing predictive accuracy and clarity. By enabling users to choose the appropriate model for a specific property, Pix2Geomodel delivers robust subsurface reservoir characterization with high visual fidelity and geological accuracy.

\begin{table}[htbp]
\centering
\caption{Key hyperparameters and training settings for the Pix2Geomodel framework.}
\label{tab:2}
\scriptsize
\begin{tabular}{|p{2.6cm}|p{4.2cm}|p{7.5cm}|}
\hline
\textbf{Category} & \textbf{Parameter} & \textbf{Value or Description} \\
\hline
\multirow{5}{*}{General settings} 
  & Image dimensions & 256 $\times$ 256 (input/output images) \\
  & Batch size & 1 (for better results in U-Net with Pix2Pix) \\
  & Buffer size & 400 (used for shuffling training data) \\
  & Random crop dimensions & 256 $\times$ 256 (cropped from 286 $\times$ 286 for data augmentation) \\
  & Normalization range & $[-1, 1]$ (input pixel values normalized to this range) \\
\hline
\multirow{6}{*}{Generator} 
  & Architecture & U-Net (with skip connections) \\
  & Activation function & Tanh (final layer) \\
  & Dropout & 50\% (applied in the first 3 upsampling layers) \\
  & Optimizer & Adam (learning rate: 2e-4, beta1: 0.5) \\
  & Loss function & GAN loss + L1 loss \\
  & L1 loss weight ($\lambda$) & 100 (balances L1 loss with GAN loss) \\
\hline
\multirow{4}{*}{Discriminator} 
  & Architecture & PatchGAN (70 $\times$ 70 patch-based discriminator) \\
  & Activation function & Leaky ReLU \\
  & Loss function & Binary Cross-Entropy Loss \\
  & Optimizer & Adam (learning rate: 2e-4, beta1: 0.5) \\
\hline
\multirow{2}{*}{Data augmentation}
  & Augmentation techniques & Random cropping, horizontal flipping, resizing \\
  & Rotation & Random (not explicitly mentioned in code but often included in augmentation pipelines) \\
\hline
\multirow{5}{*}{Training} 
  & Number of steps & 1,000 (example from the provided code; can be extended based on dataset size and desired accuracy) \\
  & Checkpoint frequency & Every 5,000 steps \\
  & Logging interval & Every 10 steps (loss values logged to TensorBoard) \\
  & Visualization interval & Every 1,000 steps (sample generated images visualized) \\
  & TensorBoard logging & Loss metrics (gen\_total\_loss, gen\_gan\_loss, gen\_l1\_loss, disc\_loss) \\
\hline
\end{tabular}
\label{tab:2}  
\end{table}

\subsection{Computing environment}
The Pix2Geomodel framework was developed and executed at the Seismic AI Lab, King Fahd University of Petroleum and Minerals, on a high-performance workstation equipped with 4$\times$ NVIDIA RTX A5500 GPUs, each with 24GB VRAM, and 128GB RAM, enabling efficient processing of large-scale datasets and complex model training. Python served as the primary programming language, leveraging libraries such as TensorFlow and PyTorch for model implementation, alongside pandas, NumPy, and Matplotlib for data manipulation, preprocessing, augmentation, and visualization, with custom scripts automating dataset preparation and splitting to ensure consistency and scalability.

\subsection{Evaluation metrics and visualization}
The Pix2Geomodel framework’s performance is comprehensively evaluated through an integrated approach combining quantitative metrics and qualitative visualizations to ensure robust assessment of its ability to generate accurate and geologically realistic subsurface reservoir property maps, emphasizing both numerical precision and geological plausibility.

\paragraph{(i) Quantitative metrics:} The evaluation employs standard semantic segmentation metrics to measure pixel-level accuracy and spatial consistency between predicted and reference maps, including pixel accuracy (PA), which calculates the proportion of correctly classified pixels; mean pixel accuracy (mPA), averaging accuracy across classes to address imbalance; mean intersection over union (mIoU), quantifying overlap between predicted and ground truth regions; and frequency-weighted IoU (FWIoU), adjusting mIoU by weighting class IoUs based on their frequency, as summarized in Table~\ref{tab:3}. Widely used in geoscience applications for balanced evaluation of imbalanced datasets, these metrics are applied post-training to validate prediction accuracy without influencing the training loss function, ensuring a balanced and representative assessment of class-specific characteristics and confirming the framework’s robustness for subsurface reservoir characterization~\cite{ioffe2015batch,ronneberger2015unet}.

\begin{table}[htbp]
\centering
\caption{Segmentation metrics and their definitions}
\label{tab:3}
\scriptsize
\begin{tabular}{|p{5.5cm}|p{8cm}|}
\hline
\textbf{Metric} & \textbf{Definition} \\
\hline
Pixel accuracy (PA) & Proportion of correctly classified pixels across the dataset. \\
\hline
Mean pixel accuracy (mPA) & Average accuracy across all classes, addressing class imbalance. \\
\hline
Mean intersection over union (mIoU) & Measures overlap between predicted and ground truth regions, widely used in semantic segmentation. \\
\hline
Frequency-weighted IoU (FWIoU) & Adjusts mIoU by weighting class IoUs according to their occurrence frequencies. \\
\hline
\end{tabular}
\end{table}

\paragraph{(ii) Qualitative visualization:} Complementing these metrics, qualitative assessments evaluate geological plausibility through visual comparisons of predicted and ground truth property maps to identify discrepancies and verify spatial continuity of geological features, discriminator heatmaps to reveal areas of high or low confidence in generated outputs, aiding in diagnosing generator weaknesses (Fig.~\ref{fig:14A}), and training progress plots of metrics like total generator loss, GAN loss, L1 loss, and discriminator loss to monitor convergence and detect overfitting or underfitting, providing visual insights into model performance.

\paragraph{(iii) Implementation and monitoring:} All metrics and visualizations are implemented using TensorBoard, enabling real-time monitoring of training progress, with logging ensuring transparency and facilitating detailed performance analysis. This combined approach validates the Pix2Geomodel framework’s effectiveness in predicting reservoir properties and provides actionable insights for further optimization (Table~\ref{tab:2}).

\section{Experimental results}

\subsection{Reservoir heterogeneity, augmented dataset validation \& spatial variability}

The Groningen gas field’s Rotliegend reservoir exhibited pronounced heterogeneity in facies, porosity, permeability, and water saturation, reflecting a complex depositional system and high reservoir quality. Facies distributions within the Slochteren formation transitioned from sandstone in the south to clay-rich sandstone and mudstone in the north. These sandstones showed high lateral continuity in the central field, grading into thinner, mudstone-interbedded layers northward. Porosity varied from 0.05 to 0.35, with peak values (0.20–0.24) in central sandstone-dominated zones and lower values (0.10–0.15) in northern mudstone-rich areas. Permeability ranged from 1 to 1000~mD, with higher values (500–1000~mD) in central aeolian-influenced sandstones and lower values (1–50~mD) in mudstone-interbedded layers, highlighting potential flow pathways and barriers. Water saturation showed higher values in northern clay-rich layers and lower values in central gas-bearing sandstones, consistent with fluid distribution trends. This spatial variability across the reservoir underscored its heterogeneous architecture and robust gas storage capacity, particularly in central high-porosity, high-permeability zones~\cite{dejager2018geology}.

The augmented dataset for the geomodel of Groningen gas field’s Rotliegend reservoir, with 2,350 images per property derived from 235 original layers, faithfully retained the original variability in facies, porosity, permeability, and water saturation, upholding the Slochteren Formation’s intricate geological framework within a 7.6 million-cell grid. Visual inspection of the augmented images confirmed consistent geological structures, such as fault-defined compartments and layered stratigraphy, mirroring the original dataset’s patterns (Fig.~\ref{fig:4}). These images preserve the shift from southern sandstone to northern clay-rich sandstone and mudstone. Permeability distributions, ranging from 1 to 1000~mD, consistently showed 500–1000~mD in central sandstone zones and 1–50~mD in mudstone-interbedded zones, maintaining flow dynamics. Porosity, spanning 0.10 to 0.24, preserved 0.20–0.30 peaks in central zones and 0.10–0.15 lows in northern mudstone areas, reflecting original depositional gradients. Water saturation trends, with elevated levels in northern clay-rich zones and reduced levels in central gas zones, were preserved, aligning with fluid patterns. This fidelity to the original dataset’s spatial contrasts underscored the augmented dataset’s capacity to reflect the reservoir’s diverse geological properties.

\begin{figure}[htbp] 
    \centering
    \includegraphics[width=0.85\linewidth]{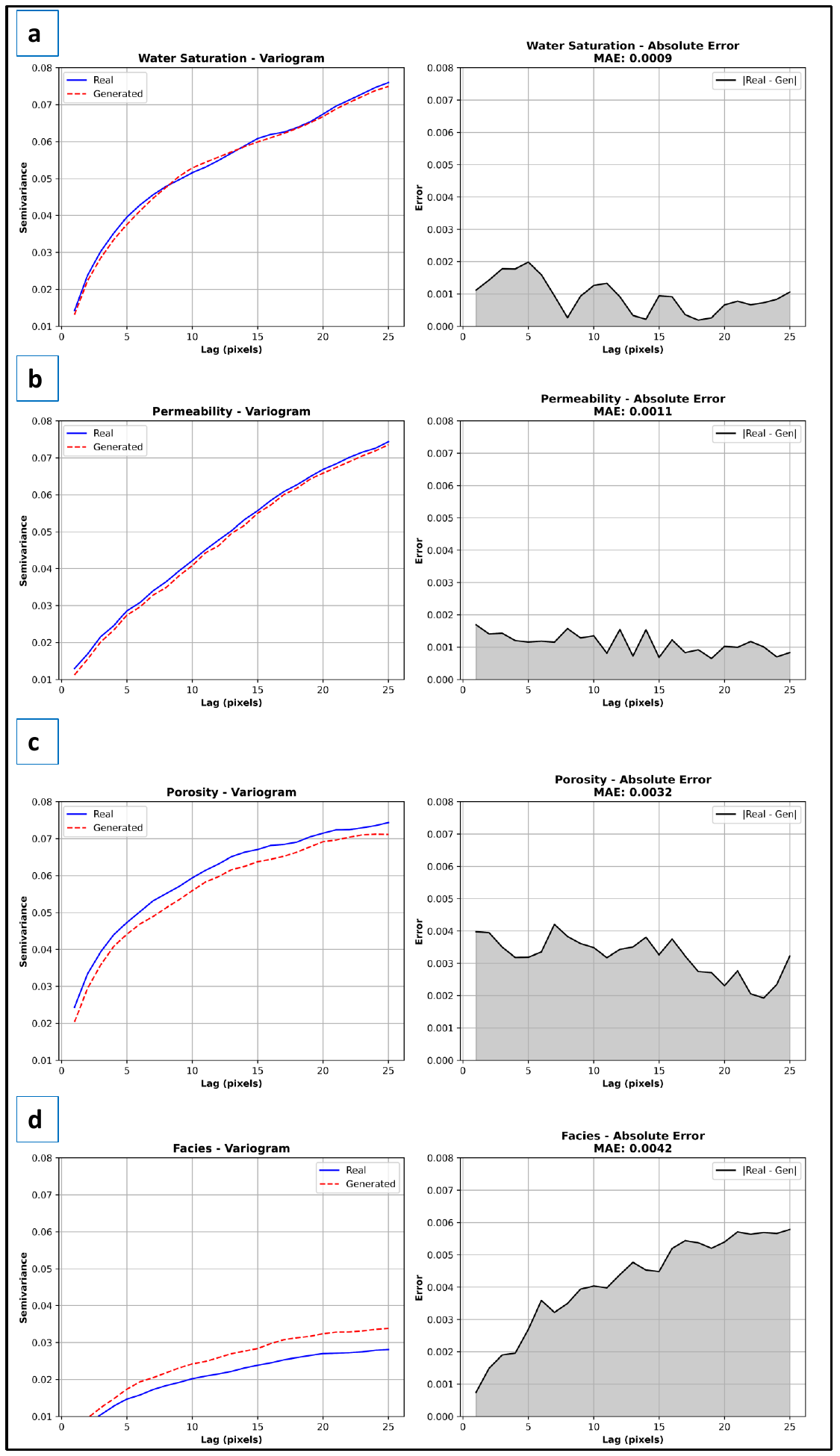}  
    \caption{Horizontal experimental variograms comparing real and Pix2Geomodel-generated spatial continuity for porosity, permeability, and facies.}
\label{fig:9}   
\end{figure}

Spatial correlation analysis of Pix2Geomodel-generated property maps for the Groningen gas field’s Rotliegend reservoir revealed a strong match with real data, as shown by horizontal variograms (Fig.~\ref{fig:9}). Variograms measure how semivariance increases with distance (lag), providing insight into spatial continuity and heterogeneity in geological data. For a given lag distance $h$, the empirical variogram $\gamma(h)$ is defined as~\cite{ag1978mining}:
\[
\gamma(h) = \frac{1}{2N(h)} \sum [ z(x_i) - z(x_i + h) ]^2 \tag{5}
\]
where $z(x_i)$ is the property value at location $x_i$, and $N(h)$ is the number of pixel pairs separated by lag distance $h$. For water saturation (Sw), semivariance increased from 0.016 at lag distance 1 to 0.073 at lag distance 25 for real data, and from 0.015 to 0.072 for generated data (Fig.~\ref{fig:9}a). Permeability variograms indicated semivariance increasing from 0.012 at lag distance 1 to 0.071 at lag distance 25 for real data, and from 0.011 to 0.070 for generated data (Fig.~\ref{fig:9}b). Porosity semivariance rose from 0.022 at lag distance 1 to 0.072 at lag distance 25 for real data, and from 0.020 to 0.070 for generated data (Fig.~\ref{fig:9}c).

Facies variograms showed semivariance growing from 0.005 at lag distance 1 to 0.028 at lag distance 25 for real data, and from 0.005 to 0.033 for generated data (Fig.~\ref{fig:9}c). Variogram-based evaluation closely matched the model’s quantitative performance metrics (Tables~\ref{tab:4}--\ref{tab:5}). Water saturation and permeability showed the strongest spatial alignment between real and generated data, with near-perfect variogram overlap and minimal error, reflected by their low mean absolute error (MAE) curves. Porosity followed closely, with slightly more deviation at short lag distances but still maintaining a low error profile. Facies exhibited the largest spatial mismatch, with a consistently higher variogram error across lags, its MAE was the highest among all properties. This is consistent with its lower class-wise performance metrics. These absolute error plots (Fig.~\ref{fig:9}, right column) provide a valuable quantitative spatial validation that complements traditional accuracy metrics.

\subsection{Model training and performance}

\begin{figure}[htbp]
    \centering
    \begin{subfigure}[t]{0.99\textwidth}
        \centering
        \includegraphics[width=\textwidth]{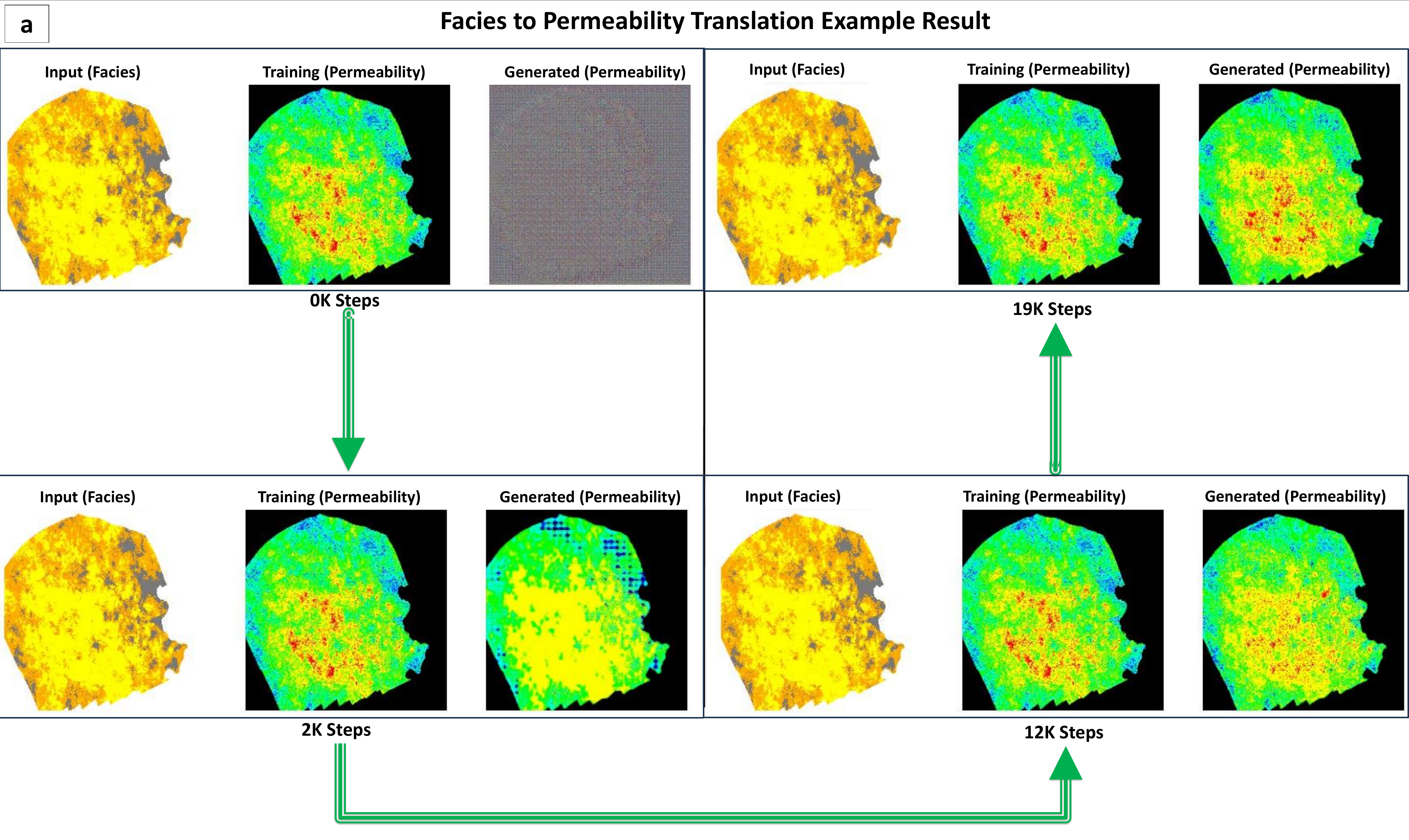}
      
        \label{fig:10a}
    \end{subfigure}
    \hfill
    \begin{subfigure}[t]{0.99\textwidth}
        \centering
        \includegraphics[width=\textwidth]{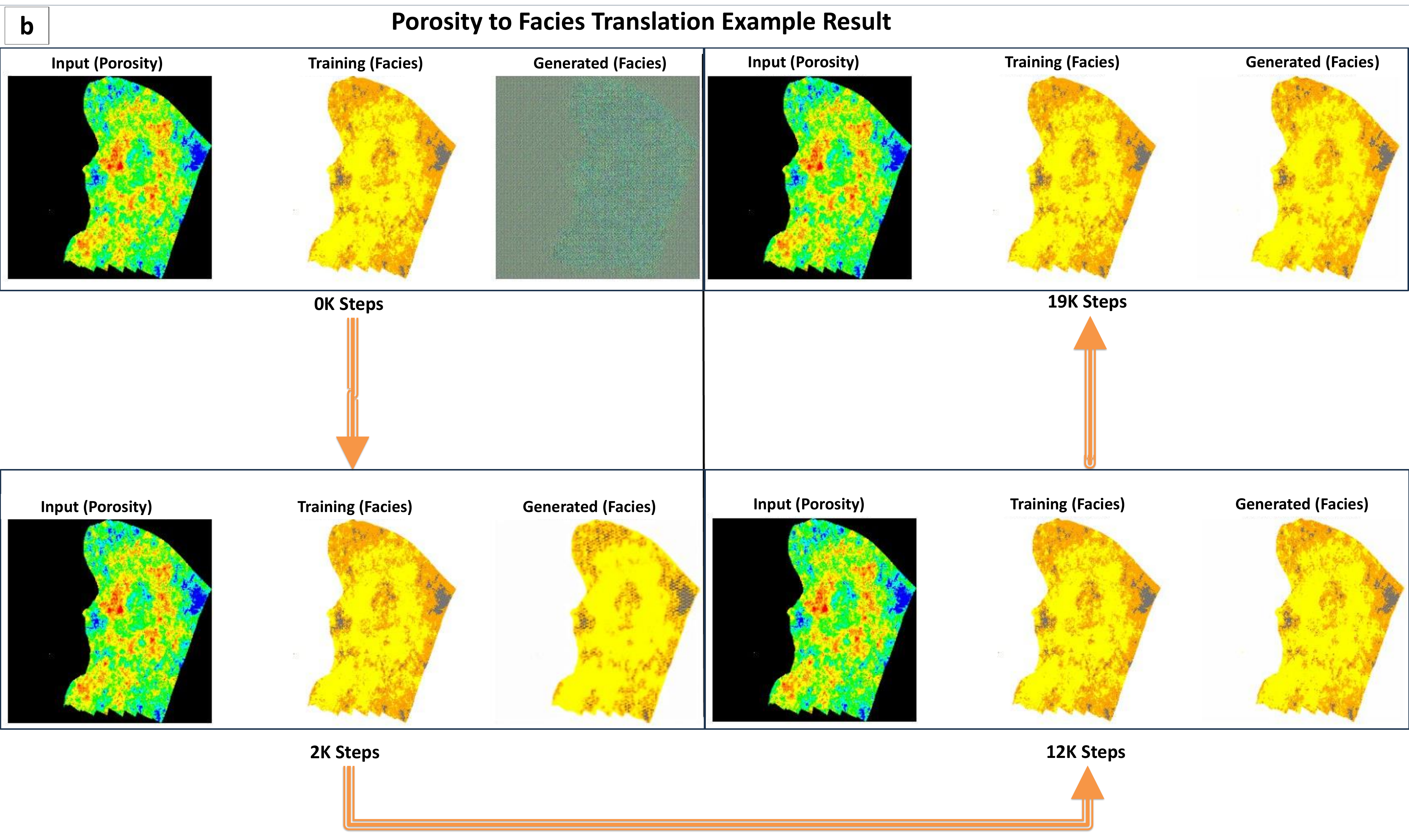}
      
        \label{fig:10b}
    \end{subfigure}
    \caption{Pix2Geomodel model’s progressive improvement in predicting permeability and facies across training steps. Input images are (facies and porosity), ground truth, and predicted outputs are shown from up-left to up-right. Predictions evolve from noise at step 0k to accurate, detailed representations by step 19k, reflecting the model’s ability to capture both large-scale and fine-grained geological features.}
    \label{fig:10}
\end{figure}

\begin{figure}[htbp] 
    \centering
    \includegraphics[width=0.75\linewidth]{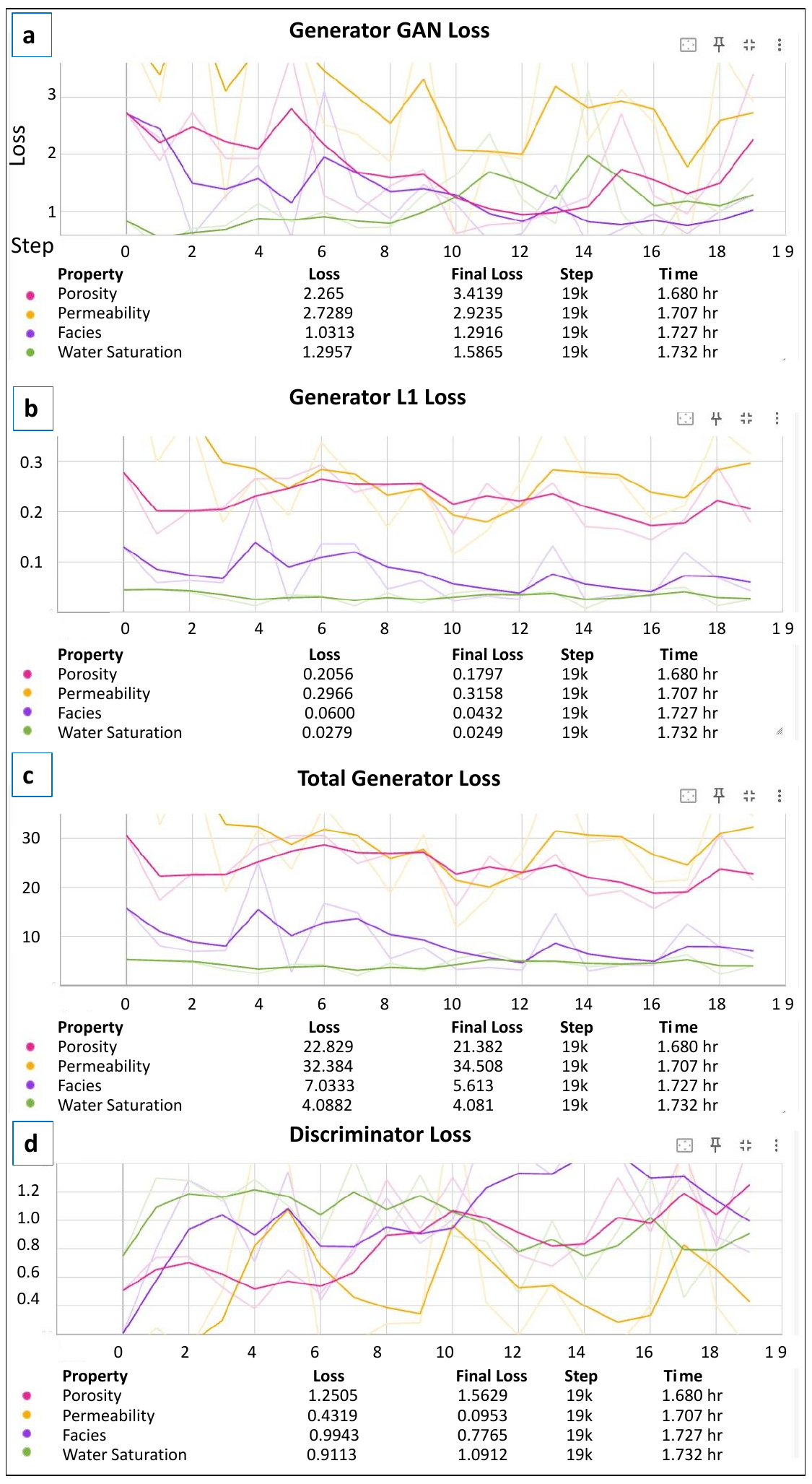}  
    \caption{Training loss curves for Pix2Geomodel: (a) Discriminator loss, (b) Generator GAN loss, (c) Generator L1 loss, and (d) Total generator loss. These curves illustrate the model’s convergence behavior across training steps, supporting the structural accuracy trends shown in Figure~\ref{fig:9}. Colored curves represent different properties: permeability (orange), porosity (pink), facies (purple), and water saturation (green). The loss trajectories reflect property-specific learning dynamics, with water saturation and facies exhibiting lower and more stable losses, while porosity and permeability show higher losses, indicating greater complexity in modeling continuous variables.}
\label{fig:11}   
\end{figure}

The Pix2Geomodel demonstrated progressive refinement in predicting reservoir properties of the Groningen gas field’s Rotliegend reservoir, including facies, porosity, permeability, and water saturation across training steps (Fig.~\ref{fig:10}), supported by qualitative visualizations and quantitative loss metrics. The evolution of permeability predictions using facies as inputs (Fig.~\ref{fig:10}a), or facies prediction using porosity as input (Fig.~\ref{fig:10}b), unfolds across panels arranged from top-left to top-right, displaying input images, ground truth, and predicted outputs at steps 0k, 1k, 2k, 12k, and 19k. At step 0k, the predicted permeability map appears as random noise. Between steps 1k and 2k, predictions begin to exhibit large-scale permeability trends, with central high-permeability zones (500–1000~mD) starting to emerge, though finer details like mudstone interbedded zones (1–50~mD) remain indistinct. After step 12k, the outputs show marked improvement, with clearer delineation of flow pathways and barriers, and by step 19k, the predictions closely mirror the ground truth, accurately capturing both broad geological patterns and fine-grained permeability variations. Concurrently, training loss curves quantify this refinement over epochs, plotted for permeability (orange), porosity (pink), facies (purple), and water saturation (green) across four subfigures (Fig.~\ref{fig:11}).

The generator GAN loss curves illustrate the adversarial learning progress of Pix2Geomodel as it attempts to produce outputs convincing enough to fool the discriminator. Early in training, all models begin with high loss values above 2.5. Porosity and permeability maintain relatively elevated and fluctuating GAN loss values, stabilizing at approximately $\sim$3.4 and $\sim$2.9, respectively, suggesting a more difficult adversarial learning process. In contrast, the generator losses for facies and water saturation decrease more consistently, stabilizing around 1.3 and 1.6, respectively (Fig.~\ref{fig:11}a).

The generator L1 loss curves provide insight into Pix2Geomodel’s accuracy in reconstructing fine-grained geological features by measuring pixel-wise differences between predicted and ground truth maps. Permeability maintains the highest L1 loss ($\sim$0.32), while porosity is slightly lower ($\sim$0.18). Facies and water saturation achieve significantly lower and more stable L1 losses ($\sim$0.04 and $\sim$0.025, respectively) (Fig.~\ref{fig:11}b).

The total generator loss combines both adversarial (GAN) and reconstruction (L1) components, providing a holistic view of Pix2Geomodel’s training convergence and predictive accuracy. Initially high for all property types, the loss trends show varying degrees of reduction over training steps. The porosity and permeability models exhibit the highest total generator loss values, stabilizing at $\sim$21.4 and $\sim$34.5, respectively, indicating persistent difficulty in balancing realism and structural accuracy for these continuous, heterogeneous properties. Conversely, the facies and water saturation models achieve significantly lower final total losses, around $\sim$5.6 and $\sim$4.1, respectively (Fig.~\ref{fig:11}c).

The discriminator loss curve reflects the Pix2Geomodel framework’s evolving capacity to distinguish real geological property maps from those generated during training. Porosity exhibits the highest final discriminator loss ($\sim$1.56), while water saturation and facies settle around $\sim$1.1 and $\sim$0.77, respectively. Permeability demonstrates the lowest final loss ($\sim$0.095), indicating that the discriminator easily distinguishes generated permeability maps from real ones (Fig.~\ref{fig:11}d).

\subsection{Masked property prediction task}

\begin{figure}[htbp]
    \centering
    \begin{subfigure}[t]{0.99\textwidth}
        \centering
        \includegraphics[width=\textwidth]{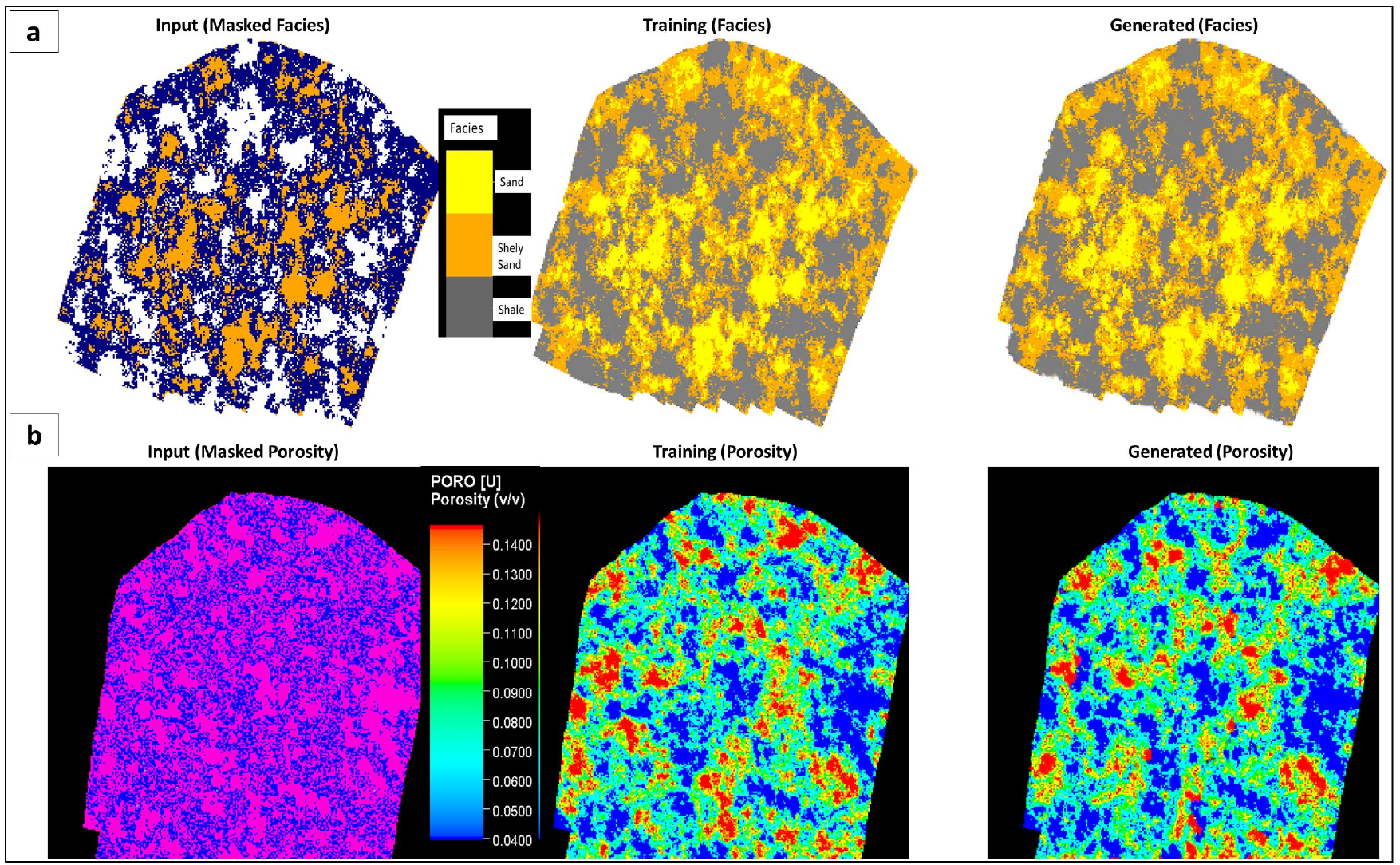}
      
        \label{fig:12a, 12b}
    \end{subfigure}
    \hfill
    \begin{subfigure}[t]{0.99\textwidth}
        \centering
        \includegraphics[width=\textwidth]{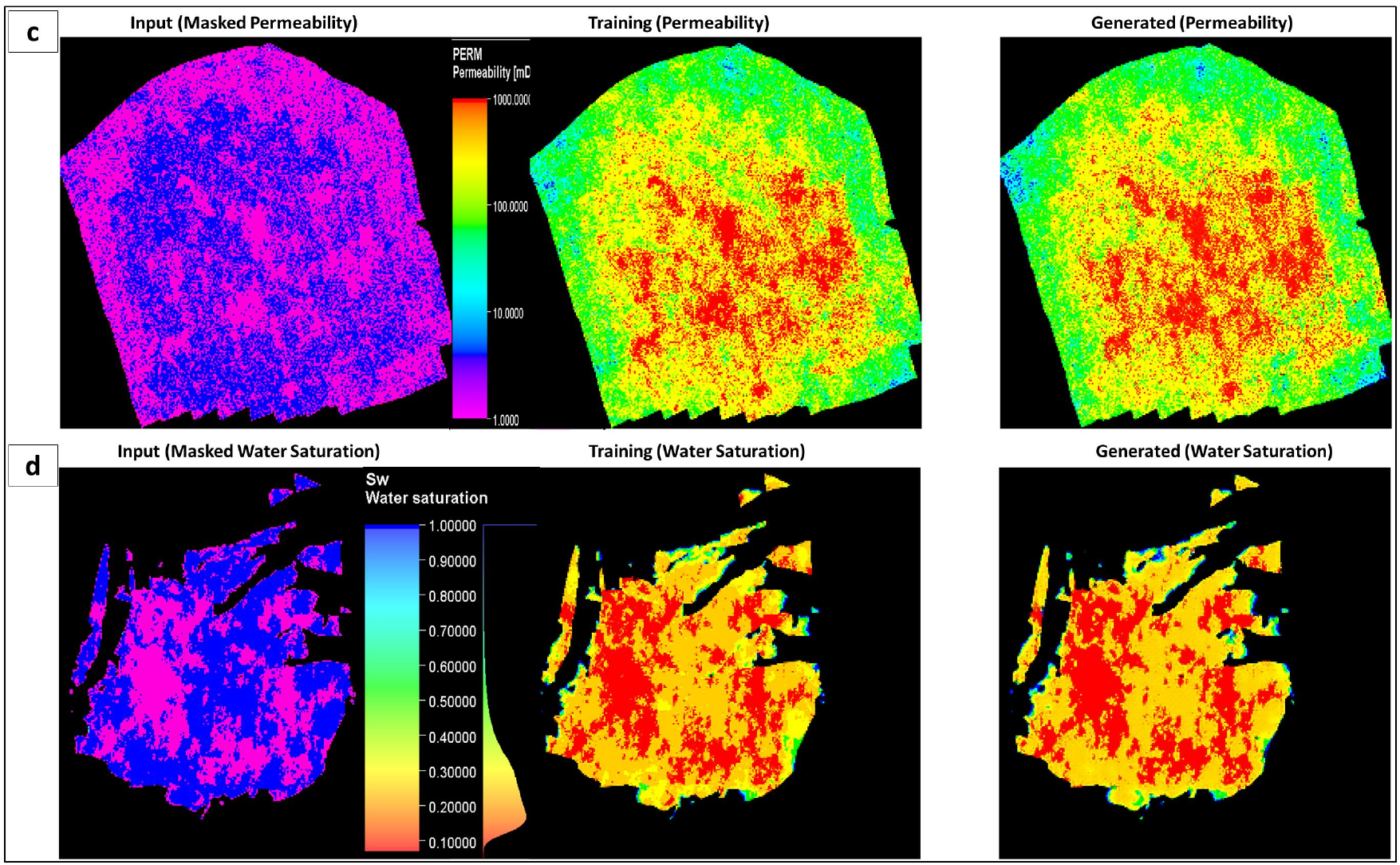}
      
        \label{fig:12c,12d}
    \end{subfigure}
    \caption{Pix2Geomodel predictions for masked inputs. (a) Facies, (b) Porosity, (c) Permeability, and (d) Water Saturation predictions compared to training. The model effectively reconstructs missing data from masked inputs, capturing spatial heterogeneity across all properties.}
    \label{fig:12}
\end{figure}

The Pix2Geomodel framework underwent rigorous evaluation for its ability to reconstruct geological properties, facies, porosity, permeability, and Sw, from masked inputs (Fig.~\ref{fig:12}a--\ref{fig:12}d, Table~\ref{tab:4}), generated through the automated annotation process outlined in Section~3.2. The framework’s adaptability shines through in capturing the intricate spatial variability highlighted in the EDA (Section~3.1, Fig.~\ref{fig:3}), where property distributions revealed significant heterogeneity across the 7.6 million-cell grid.

\begin{table}[htbp]
\centering
\caption{Best results for each property prediction task}
\label{tab:4}
\begin{tabular}{|c|c|c|c|c|}
\hline
{Metric} & {Facies} & {Porosity} & {Permeability} & {Water Saturation} \\
\hline
{PA} & \textbf{0.88} & 0.70 & 0.74 & \textbf{0.96} \\
{mPA} & 0.36 & \textbf{0.66} & 0.51 & 0.46 \\
{Mean IoU} & 0.31 & \textbf{0.50} & 0.39 & 0.37 \\
{FWIoU} & \textbf{0.85} & 0.55 & 0.60 & \textbf{0.95} \\
\hline
\end{tabular}
\label{tab:5}
\end{table}

The performance of Pix2Geomodel in predicting facies and porosity from masked inputs serves as a robust testament to its ability to reconstruct geologically realistic property maps under data-scarce conditions. For facies prediction, the model achieved a PA of 0.88, a mean mPA of 0.3585, a mean IoU of 0.3121, and a FWIoU of 0.85 (Fig.~\ref{fig:12}a, Table~\ref{tab:4}). For porosity prediction, the model recorded a PA of 0.70, an mPA of 0.6598, a mean IoU of 0.4956, and an FWIoU of 0.55, successfully replicating both small-scale fluctuations and large-scale trends within the reservoir (Fig.~\ref{fig:12}b, Table~\ref{tab:4}).

Pix2Geomodel’s predictive performance for permeability and water saturation under masked input conditions further demonstrates its capability to reconstruct geologically realistic property maps under data-scarce conditions. For permeability prediction, the model attained a PA of 0.74, an mPA of 0.5121, a mean IoU of 0.3948, and an FWIoU of 0.60 (Fig.~\ref{fig:12}c, Table~\ref{tab:4}). For water saturation prediction, the model achieved a PA of 0.96, an mPA of 0.4579, a mean IoU of 0.3710, and FWIoU of 0.95 (Fig.~\ref{fig:12}d, Table~\ref{tab:4}).

\subsection{Property-to-property translation task}

\begin{figure}[htbp]
    \centering
    \begin{subfigure}[t]{0.99\textwidth}
        \centering
        \includegraphics[width=\textwidth]{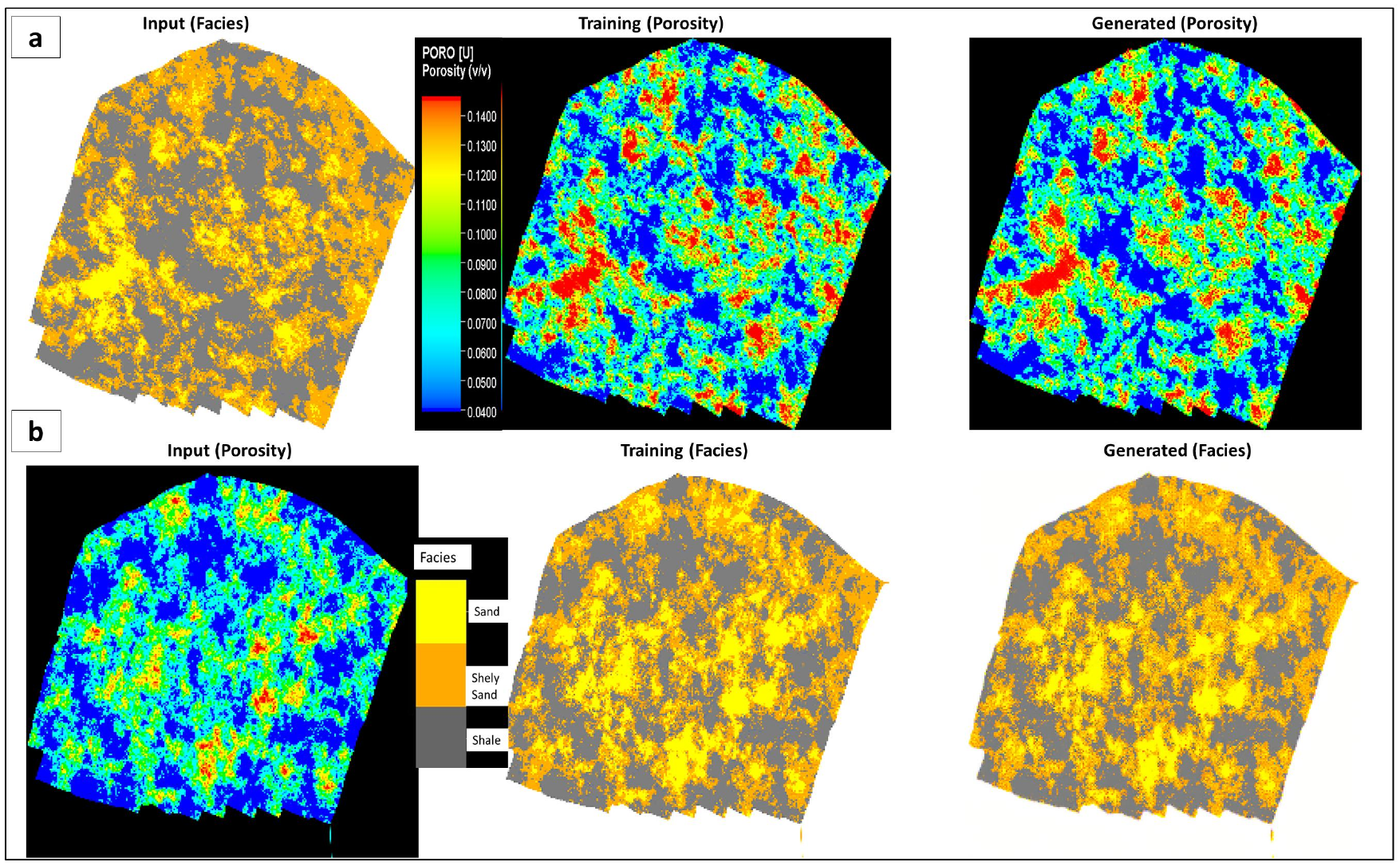}
      
        \label{fig:13a, 13b}
    \end{subfigure}
    \hfill
    \begin{subfigure}[t]{0.99\textwidth}
        \centering
        \includegraphics[width=\textwidth]{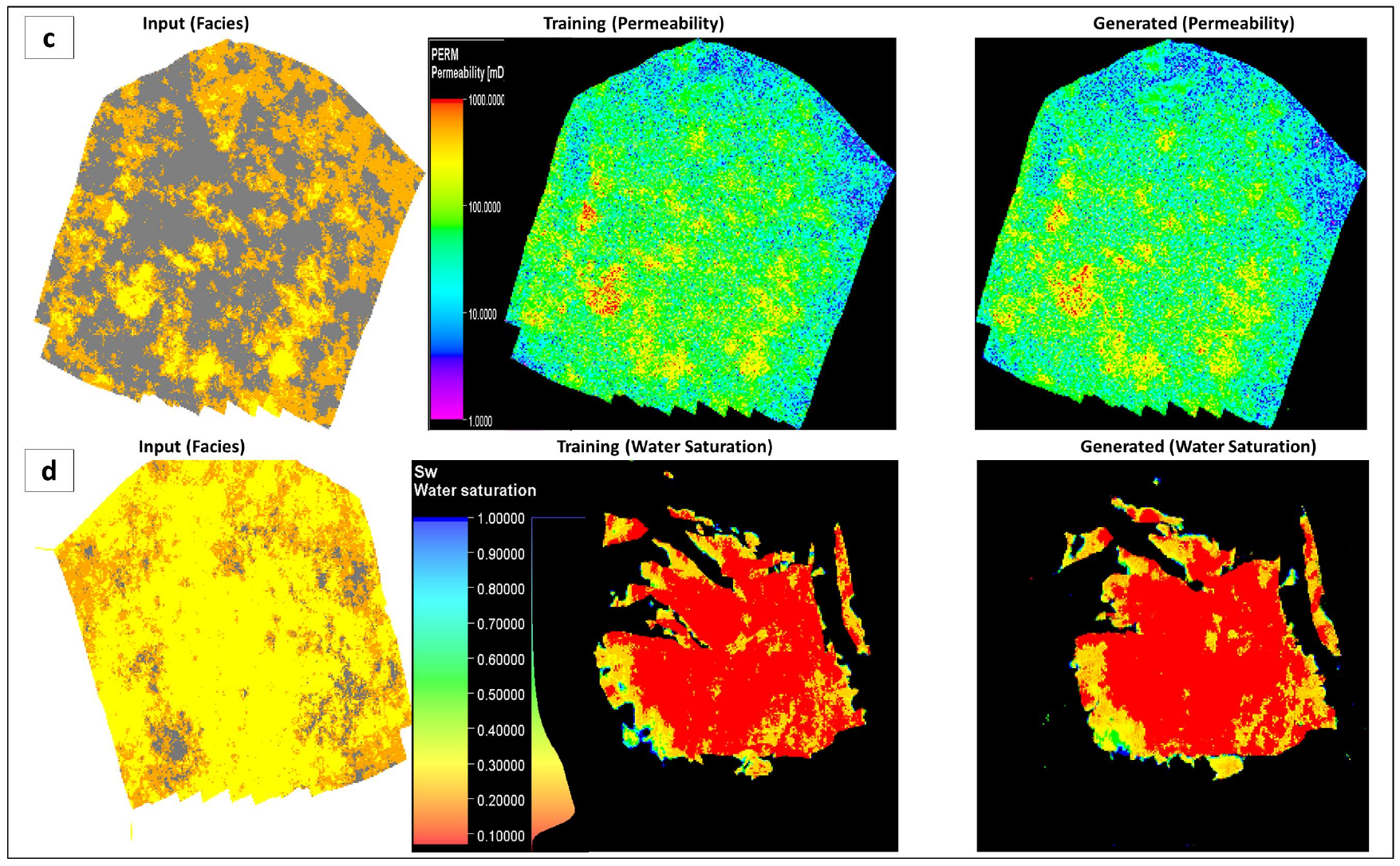}
      
        \label{fig:13c,13d}
    \end{subfigure}
    \caption{Property-to-property translation using Pix2Geomodel. (a) Facies-to-porosity; (b) Porosity-to-facies; (c) Facies-to-permeability; (d) Facies-to-water saturation. Each triplet shows input (left), training (center), and generated output (right).}
    \label{fig:13}
\end{figure}

The efficacy of the Pix2Geomodel framework in translating between different geological properties was rigorously assessed under the property-to-property translation task, aiming to determine its capability to predict one geological attribute from the input of another, a key aspect for comprehensive reservoir characterization. These outcomes focus on four distinct translation scenarios: facies-to-porosity, porosity-to-facies, facies-to-permeability, and facies-to-Sw (Table~\ref{tab:5}, Fig.~\ref{fig:13}a--\ref{fig:13}d).

\begin{table}[htbp]
\centering
\caption{Best results for each property-to-property translation task}
\label{tab:5}
\begin{tabular}{|c|c|c|c|c|c|}
\hline
{Property to property} & {PA} & {mPA} & {Mean IoU} & {FWIoU} \\
\hline
Facies to Porosity & 0.58 & 0.5831 & 0.3930 & 0.41 \\
Porosity to Facies & \textbf{0.94} & 0.3961 & 0.3413 & \textbf{0.91} \\
Facies to Permeability & 0.70 & 0.4673 & 0.3632 & 0.56 \\
Facies to Sw & \textbf{0.98} & 0.3328 & 0.3278 & \textbf{0.97} \\
\hline
\end{tabular}
\label{tab:6}
\end{table}

In the facies-to-porosity translation scenario, Pix2Geomodel generated porosity distributions from facies data, where the model captured both overarching patterns and nuanced, localized variations in porosity, achieving a PA of 0.58, a mPA of 0.5831, a mean IoU of 0.3930, and a FWIoU of 0.41 (Fig.~\ref{fig:13}a, Table~\ref{tab:5}). In the porosity-to-facies translation scenario, Pix2Geomodel tackled the challenge of converting continuous porosity values into discrete facies distributions, retaining significant facies boundaries and internal heterogeneities. The predicted facies maps aligned notably well with the reference data, with a PA of 0.94, an mPA of 0.3961, a mean IoU of 0.3413, and an FWIoU of 0.91 (Fig.~\ref{fig:13}b, Table~\ref{tab:5}). In the facies-to-permeability translation scenario, Pix2Geomodel predicted permeability from facies data, with the resulting maps showing a strong correlation with the ground truth test dataset. The model captured broad trends and finer-scale heterogeneities, recording a PA of 0.70, an mPA of 0.4673, a mean IoU of 0.3632, and an FWIoU of 0.56 (Fig.~\ref{fig:13}c, Table~\ref{tab:5}). In the final facies-to-Sw translation scenario, the framework achieved a PA of 0.98, an mPA of 0.3328, a mean IoU of 0.3278, and an FWIoU of 0.97 (Fig.~\ref{fig:13}d, Table~\ref{tab:5}).

\section{Discussion}

The findings from Sections~3.1 to 3.4 provide a comprehensive view of the Pix2Geomodel framework’s performance in addressing the challenges of reservoir characterization within the Groningen gas field’s Rotliegend reservoir. This discussion interprets the implications of these results, evaluates the framework’s strengths and limitations, compares its performance with prior generative AI approaches in geoscience, and explores avenues for future research. By focusing on the spatial variability, training dynamics, masked prediction, and property translation tasks, this section underscores Pix2Geomodel’s transformative potential while identifying areas for enhancement in subsurface modeling.

\subsection{Spatial variability and dataset fidelity in reservoir characterization}

The reservoir heterogeneity analysis (Section~3.1) highlights the complex geological framework of the Groningen gas field, characterized by a transition from sandstone-dominated zones in the south to mudstone-rich layers in the north, with distinct property variations across facies, porosity, permeability, and water saturation. This variability, rooted in the depositional and diagenetic history of the Slochteren formation, poses significant challenges for traditional geomodeling methods, as noted in the introduction section. The observed property ranges, porosity from 0.05 to 0.35, permeability from 1 to 1000~mD, and the inferred water saturation trends, align with geological expectations for the Rotliegend reservoir~\cite{dejager2018geology}. These findings emphasize the need for a modeling approach capable of capturing both large-scale trends and localized heterogeneities, a challenge that Pix2Geomodel addresses through its conditional GAN framework.

The augmented dataset’s fidelity to the original 235-layer dataset, with 2,350 images per property, is a critical strength. By preserving the spatial, the augmented dataset ensures that Pix2Geomodel’s training inputs reflect the reservoir’s true geological complexity. This fidelity, supported by visual inspections (Section~2.1, Fig.~\ref{fig:4}), aligns with the methodology’s emphasis on robust data augmentation (Section~2.2), which reduces overfitting while maintaining geological realism. The spatial correlation analysis via variograms (Section~3.1, Fig.~\ref{fig:9}) further validates this approach, as the close match between real and generated semivariance trends (e.g., porosity semivariance rising from 0.02 to 0.07 for real data versus 0.022 to 0.072 for generated data, accompanied by small absolute errors typically below 0.005) indicates that Pix2Geomodel captures the spatial continuity and heterogeneity inherent in the reservoir. This capability is particularly valuable for applications requiring accurate representation of flow pathways and barriers, such as hydrocarbon recovery and reservoir simulation, aligning with the study’s broader objectives (Introduction Section).

\subsection{Training dynamics and model convergence}

The training dynamics of Pix2Geomodel, as analyzed in Section~3.2, reveal the framework’s ability to iteratively refine its predictions across facies, porosity, permeability, and water saturation. The progressive improvement in permeability predictions, from noise-like outputs at step 0k to high-fidelity maps by step 19k (Fig.~\ref{fig:10}a), underscores the effectiveness of the adversarial training process outlined in the methodology (Section~2.3). The ability to delineate central high-permeability zones (500–1000~mD) and interbedded mudstone areas (1–50~mD) by the final training steps suggests that Pix2Geomodel successfully learns the spatial relationships important for reservoir flow modeling. This iterative learning process is a key advantage over traditional geostatistical methods, which often struggle with such dynamic variability (Introduction Section).

The loss curves (Fig.~\ref{fig:11}a--d) provide further insight into Pix2Geomodel’s convergence behavior. The smoother convergence of discriminator loss (Fig.~\ref{fig:11}d) for facies and porosity, with final discriminator loss values around 0.78 and 1.56, respectively, compared to the more variable and significantly lower final loss for permeability (0.095) and relatively stable final loss for water saturation (1.09), suggests that the model finds it easier to consistently distinguish real versus generated maps for facies. Permeability, with its more pronounced final discriminator loss drop to 0.095, underscores an effective discriminator performance in recognizing permeability features, likely reflecting strong spatial heterogeneity.

Generator GAN losses (Fig.~\ref{fig:11}a) further illustrate this challenge, showing higher and more variable final losses for permeability (2.92) and water saturation (1.59), reflecting their broader value ranges and complex distributions. The total generator loss (Fig.~\ref{fig:11}c) reinforces this complexity, with permeability showing significantly higher values (34.51) compared to porosity (21.38), indicating permeability’s greater prediction difficulty, likely influenced by multiscale geological processes such as diagenesis and fracturing~\cite{dejager2018geology}.

The lower generator L1 losses (Fig.~\ref{fig:11}b) for facies (0.043) and water saturation (0.025) compared to higher values for permeability (0.316) and porosity (0.180) emphasize better pixel-wise accuracy for facies and water saturation, likely due to their more defined and less heterogeneous spatial distributions.

These training dynamics highlight the importance of the U-Net generator and PatchGAN discriminator architecture (Section~2.3) in balancing realism and structural accuracy. However, persistent challenges in optimizing permeability predictions, highlighted by its comparatively high total generator and GAN losses, suggest that future iterations of Pix2Geomodel could benefit from specialized loss functions or multi-resolution approaches to better capture microstructural variability, as discussed later.

\subsection{Masked property prediction: balancing accuracy and geological realism}

The masked property prediction task (Section~3.3) evaluates Pix2Geomodel’s ability to reconstruct geological properties from limited input data, a critical capability for scenarios with incomplete datasets, as highlighted in the Introduction. The performance across facies, porosity, permeability, and water saturation demonstrates the framework’s adaptability to diverse property types, each with distinct spatial characteristics.

For facies and porosity predictions, the high PA (0.88 for facies, 0.70 for porosity) and FWIoU (0.85 for facies, 0.55 for porosity) indicate that Pix2Geomodel effectively captures both large-scale trends and localized variations (Fig.~\ref{fig:12}a, Fig.~\ref{Fig.15A}, Table~\ref{tab:4}). The model’s success in reconstructing facies aligns with the reservoir’s sandstone-dominated nature (80\% prevalence), where clear lithological boundaries facilitate accurate predictions. Porosity predictions, spanning the 0.10–0.24 range (Fig.~\ref{fig:12}b, Fig.~\ref{Fig.15A}a, Fig.~\ref{Fig.15A}b, Table~\ref{tab:4}), reflect the model’s ability to handle the proximal-to-distal gradient noted in Section~4.1, though the lower FWIoU suggests challenges in capturing the full spectrum of small-scale fluctuations, particularly in mudstone-rich northern areas.

Permeability and water saturation predictions further showcase Pix2Geomodel’s strengths and limitations. The PA of 0.74 and FWIoU of 0.60 for permeability indicate robust performance in tracing wide-ranging values (1–1000~mD), but the mean IoU of 0.3948 (Fig.~\ref{fig:12}c, Fig.~\ref{Fig.16A}a, Table~\ref{tab:4}) suggests that finer-scale heterogeneities, such as those in peripheral zones, are less precisely captured. This is likely due to the complex interplay of depositional and diagenetic processes affecting permeability, which may require additional input data (e.g., seismic attributes) to fully resolve. Water saturation predictions, with an impressive PA of 0.96 and FWIoU of 0.95 (Fig.~\ref{fig:12}d, Fig.~\ref{Fig.16A}b, Table~\ref{tab:4}), benefit from the clearer relationship between facies boundaries and fluid distribution, as well as the rapid convergence observed in training (Section~3.2). The model’s ability to reflect the bimodal distribution (peaks at 0.2 and 0.7) underscores its potential for fluid dynamics modeling.

Overall, the masked prediction task highlights Pix2Geomodel’s ability to balance geological realism with predictive accuracy under data-scarce conditions. The framework’s performance aligns with its design goals (Introduction Section) but also reveals areas for improvement, particularly in handling properties with high microstructural variability.

\subsection{Property-to-property translation: bridging discrete and continuous domains}

The property-to-property translation task (Section~3.4) represents a significant advancement in reservoir characterization, as it enables direct mapping between geological properties, addressing the need for integrated subsurface modeling (Introduction Section). The four scenarios, facies-to-porosity, porosity-to-facies, facies-to-permeability, and facies-to-water saturation, demonstrate Pix2Geomodel’s flexibility and robustness across diverse translation challenges.

The facies-to-porosity translation, with a PA of 0.58 and FWIoU of 0.41 (Fig.~\ref{fig:13}a, Fig.~\ref{Fig.17A}a, Table~\ref{tab:5}), reflects the model’s ability to capture overarching porosity patterns while struggling with localized variations. This performance is understandable given the complex relationship between facies and porosity, where depositional environments and diagenetic overprints introduce non-linear dependencies~\cite{dejager2018geology}. The higher mPA (0.5831) suggests that the model performs better on a per-class basis, but the lower FWIoU indicates that rare classes (e.g., low-porosity mudstones) may be underrepresented in predictions.

In contrast, the porosity-to-facies translation achieves a higher PA (0.94) and FWIoU (0.91) (Fig.~\ref{fig:13}b, Fig.~\ref{Fig.17A}b, Table~\ref{tab:5}), indicating strong performance in bridging continuous and discrete domains. The model’s success in retaining facies boundaries and internal heterogeneities aligns with the reservoir’s sandstone dominance, where clear lithological transitions facilitate accurate mapping. This bidirectional translation capability is a key strength, as it enables Pix2Geomodel to support rock type distribution mapping, an indispensable aspect of reservoir characterization (Introduction Section).

The facies-to-permeability translation, with a PA of 0.70 and FWIoU of 0.56 (Fig.~\ref{fig:13}c, Fig.~\ref{Fig.18A}a, Table~\ref{tab:5}), highlights Pix2Geomodel’s utility in modeling flow characteristics. The model’s ability to capture broad trends and finer-scale heterogeneities aligns with the reservoir’s permeability gradients (500–1000~mD in central zones versus 1–50~mD in peripheral areas). However, the mean IoU of 0.3632 suggests that the model may struggle with the full range of permeability variations, particularly in areas influenced by fracturing or cementation, which introduce additional complexity.

The facies-to-water saturation translation stands out with a PA of 0.98 and FWIoU of 0.97 (Fig.~\ref{fig:13}d, Fig.~\ref{Fig.18A}b, Table~\ref{tab:5}), reflecting the model’s proficiency in fluid-distribution modeling. The strong correlation between facies boundaries and water saturation patterns, as noted in Section~3.1, likely contributes to this success, as does the rapid convergence observed during training (Section~3.2). This performance underscores Pix2Geomodel’s potential for applications requiring precise fluid dynamics modeling, such as enhanced recovery strategies (Section~4.7).

\subsection{Comparison with prior GAN-based approaches}

Pix2Geomodel builds on and extends prior GAN-based approaches in reservoir modeling, offering a novel application of the Pix2Pix framework to direct property-to-property translation. Compared to GANSim~\cite{song2021gansim,song2022gansim3d,song2023gansim}, which focused on stochastic facies generation, Pix2Geomodel’s deterministic mapping approach provides greater control over property predictions, making it more suitable for tasks requiring high fidelity to ground truth data. The framework also extends stochastic Pix2Pix applications for channelized reservoirs~\cite{pan2021stochastic} by applying the methodology to a broader range of properties, including continuous variables like porosity and permeability. Additionally, Pix2Geomodel’s performance aligns with 3D-Pix2Pix applications for history matching~\cite{luy2021pix2pix}, confirming the scalability of Pix2Pix in geoscience contexts.

Insights from related domains, such as medical imaging and cartography, further contextualize Pix2Geomodel’s performance. In medical imaging, tasks with strong physical correlations (e.g., T1-to-T2 MRI translation) often achieve high accuracy~\cite{chen2021bridging,aljohani2022generating}, similar to Pix2Geomodel’s facies-to-water saturation translation. Conversely, tasks requiring microstructural detail, such as lesion segmentation, face challenges akin to those in porosity and permeability predictions~\cite{li2024mapping}. These parallels suggest that Pix2Geomodel’s strengths lie in leveraging clear geological correlations, while its limitations in capturing fine-scale variability may require architectural enhancements, as discussed below.

\subsection{Limitations and challenges}

Despite its successes, Pix2Geomodel faces several limitations that warrant consideration. First, the framework’s performance is heavily influenced by the quality and representativeness of the training data. While the augmented dataset preserves the Groningen reservoir’s variability, it may not fully capture sub-resolution geological features, such as microfractures or thin laminations, which are extremely important for accurate permeability modeling. This limitation aligns with challenges noted in prior studies~\cite{bai2020hybrid,alqahtani2021applications}, where non-linear geological processes complicate predictions.

Second, the current 2D implementation, while computationally efficient, limits Pix2Geomodel’s ability to capture vertical continuity in the reservoir. Preliminary experiments with a 3D version (Pix2Geomodel v2.0) show promise, but computational constraints and data availability have delayed full validation. This limitation underscores the need for scalable architectures and larger datasets to support 3D modeling, a direction for future research.

Third, the disparity in performance across properties, particularly the lower accuracy for porosity and permeability compared to facies and water saturation, suggests that Pix2Geomodel may struggle with properties exhibiting high microstructural variability. The U-Net skip connections and PatchGAN discriminator enhance spatial fidelity~\cite{ronneberger2015unet,isola2017image}, but they may not fully address the multiscale nature of these properties, as noted in prior GAN applications~\cite{chen2022modeling}.

\subsection{Future directions and recommendations}

The results and limitations of Pix2Geomodel point to several avenues for future research. First, integrating multi-modal data, such as seismic attributes, borehole imaging logs, or core data, could enrich the feature representation and improve predictions for properties like permeability and porosity. Incorporating physics-based constraints, such as those proposed by Raissi et al.~\cite{raissi2019physics}, could further enhance geological realism by ensuring that predictions adhere to physical principles like Darcy’s law for fluid flow.

Second, exploring advanced GAN architectures, such as StyleGAN~\cite{karras2021style} or cycle-consistent GANs~\cite{zhu2017unpaired}, could improve the framework’s ability to capture fine-scale variability and ensure consistency in bidirectional translations. Multi-resolution approaches, which model properties at different scales, may also address the challenges of microstructural detail, as suggested by Wang et al.~\cite{wang2023scientific}.

Third, the development of Pix2Geomodel v2.0 for 3D reservoir modeling should be prioritized. A fully validated 3D implementation could enhance volumetric reservoir characterization, geomodeling, and subsurface flow simulations, aligning with the study’s long-term goals (Section~4.6). This would require addressing computational constraints through optimized architectures or distributed computing frameworks.

Finally, hybrid ML models that combine data-driven methods with physics-based modeling could bridge the gap between geological realism and predictive accuracy~\cite{degen2023perspectives}. Such approaches, as demonstrated by Fossum et al.~\cite{fossum2024ensemble} in history matching, could improve resource estimation and data integration, further expanding Pix2Geomodel’s applicability in reservoir management.

\subsection{Implications for reservoir modeling}

Pix2Geomodel’s success in masked prediction and property translation tasks underscores the potential of conditional GANs in reservoir modeling. By enabling direct property mapping and reconstruction under data-scarce conditions, the framework addresses key challenges outlined in the Introduction, such as capturing complex spatial relationships and reducing uncertainties in volumetric estimates. Its ability to model flow characteristics (facies-to-permeability) and fluid distributions (facies-to-water saturation) makes it a valuable tool for enhancing recovery strategies.

Moreover, Pix2Geomodel’s performance confirms the scalability of Pix2Pix-based approaches in geoscience, building on prior work~\cite{song2021gansim,luy2021pix2pix} and opening new avenues for predictive modeling in hydrocarbon recovery, geothermal energy, and carbon sequestration. The framework’s commitment to open science, with publicly available datasets and source code (Section~4.6), fosters reproducibility and collaboration, further amplifying its impact in the geoscience community.

\section{Conclusion}
{This study presented Pix2Geomodel, a pioneering conditional GAN framework, to enhance geological modeling of the Groningen gas field’s Rotliegend reservoir. By leveraging Pix2Pix architecture, the framework successfully predicted facies, porosity, permeability, and water saturation from masked inputs and facilitated property-to-property translation, achieving high accuracies (e.g., facies PA 0.88, water saturation PA 0.96) and robust translation performance (e.g., facies-to-Sw PA 0.98). The approach captured the reservoir’s spatial heterogeneity, validated by variogram analysis, and outperformed traditional methods in handling complex subsurface patterns. Despite challenges with porosity and permeability predictions due to microstructural variability, Pix2Geomodel demonstrated significant potential for reservoir characterization under data-scarce conditions. The study’s open-source datasets and code foster reproducibility, aligning with geoscience community goals. Future work will explore 3D modeling (Pix2Geomodel v2.0), multi-modal data integration, and advanced GAN architectures to address limitations and enhance applications in hydrocarbon recovery, geothermal energy, and carbon sequestration.}

\section*{CRediT authorship contribution statement}
\textbf{Abdulrahman Al-Fakih}: Conceptualization, Methodology, Formal Analysis, Software Development, Data Curation, Visualization, Writing Original Draft Preparation. 
\textbf{Ardiansyah Koeshidayatullah}: Resources, Supervision, Writing - Review \& Editing.
\textbf{Nabil A. Saraih}: Visualization Enhancements, Manuscript Restructuring, Writing, Review \& Editing.
\textbf{Tapan Mukerji}: Conceptual Review, Scientific Oversight, Review \& Editing.
\textbf{Rayan Kanfar}: Domain Industrial Expertise, Methodology and Results Analysis, Review \& Editing.
\textbf{Abdulmohsen Alali}: Domain Industrial Expertise, Review \& Editing.
\textbf{SanLinn I. Kaka}: Supervision, Project Administration, Review \& Editing.

\subsection*{Competing interests}
The authors declare no competing interests.

\section*{Declaration of competing interest and use of generative AI}
The authors affirm that they have no known competing financial interests or personal relationships that may have influenced the work presented in this paper. 
During the preparation of this work, the author(s) used the ChatGPT language model from OpenAI to refine grammar and enhance text coherence in this article. After using this tool/service, the author(s) reviewed and edited the content as needed and take(s) full responsibility for the content of the publication.

\section*{Data availability}
Supplementary material related to this article is available online at \url{https://github.com/ARhaman}.

Correspondence and requests for materials should be addressed to A.F. at \href{mailto:alfakihabdulrahman2030@gmail.com}{alfakihabdulrahman2030@gmail.com}.

\section*{Acknowledgements}
The authors gratefully acknowledge the College of Petroleum Engineering and Geosciences at King Fahd University of Petroleum and Minerals (KFUPM) for their generous support and facilitation of this research. Their assistance was instrumental in enabling the presentation of this work at international scientific conferences and promoting collaborative research excellence.



\section*{Appendix}

\renewcommand{\thefigure}{\arabic{figure}A} 

\begin{figure}[htbp]
    \centering
    \includegraphics[width=1\textwidth]{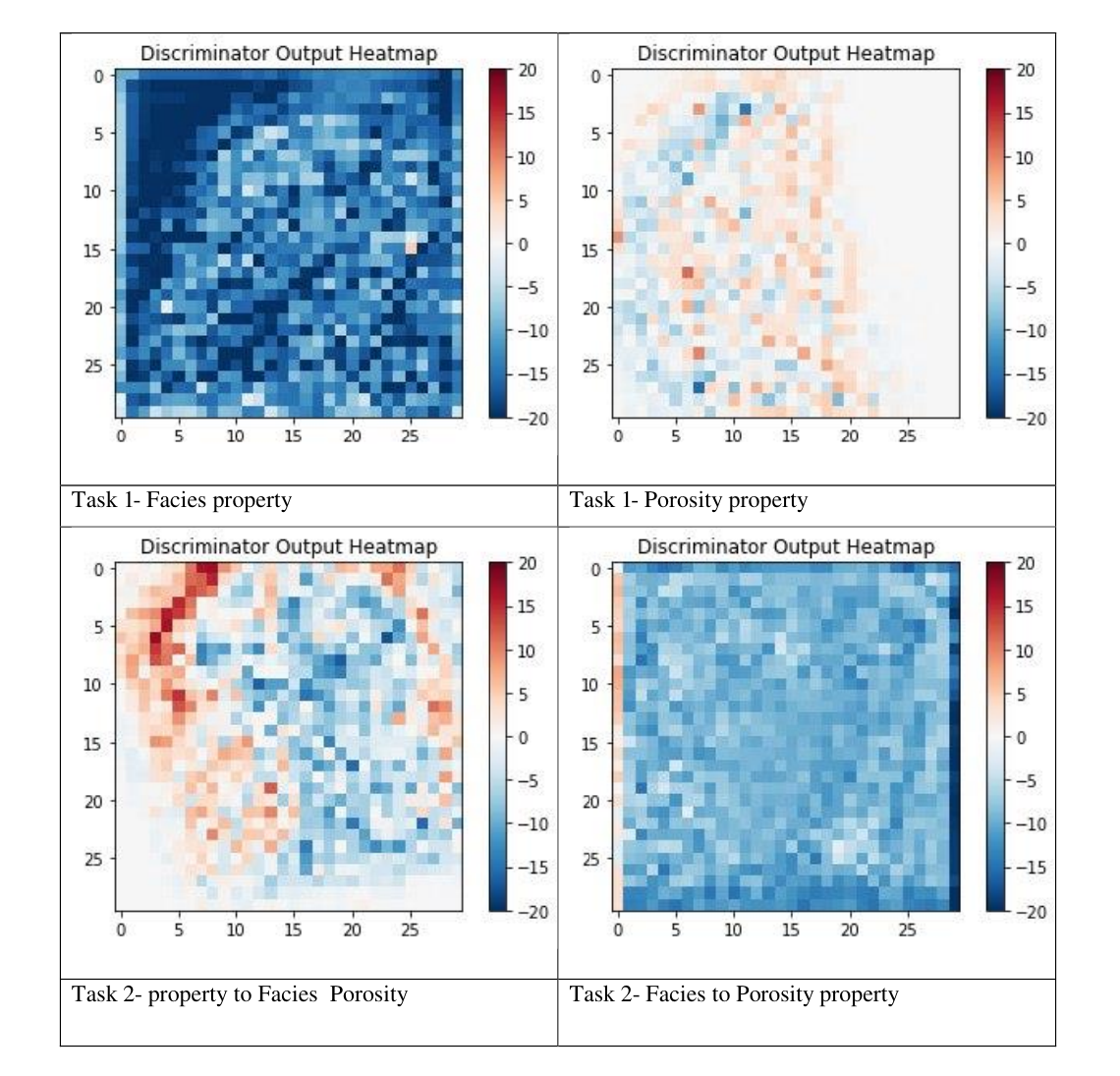}  
    \caption{Discriminator output heatmaps for Task 1 (facies and porosity properties) and Task 2 (facies-to-porosity transformations). Color gradients indicate discriminator confidence, with warmer colors reflecting higher prediction accuracy.}
    \label{Fig.14A}
\end{figure}

\begin{figure}[htbp]
    \centering
    \includegraphics[width=1\textwidth]{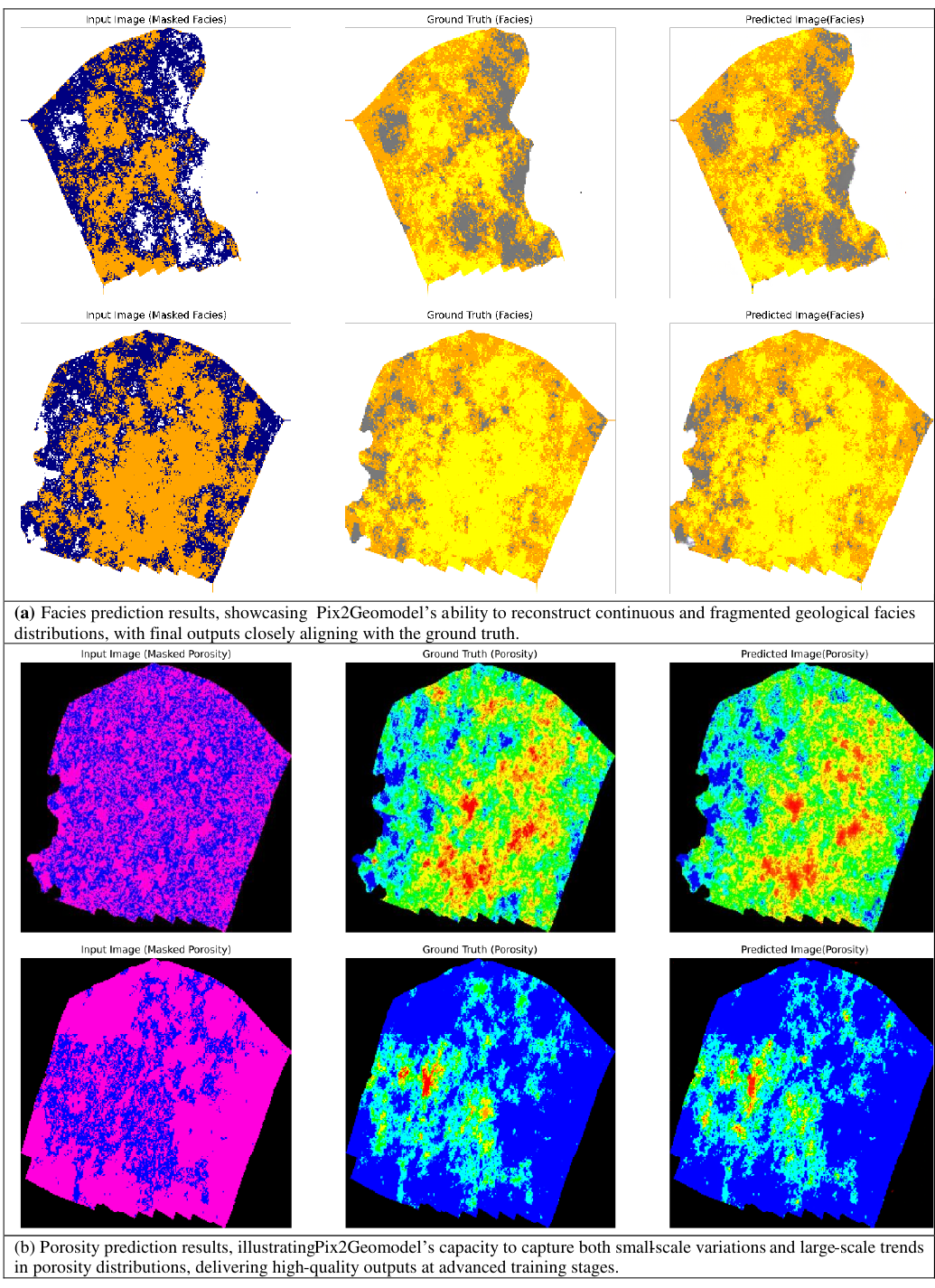}  
    \caption{Comparison of Pix2Geomodel predictions with ground truth for facies (a) and porosity (b). The progression from masked inputs to predicted outputs highlights Pix2Geomodel's ability to learn and replicate complex geological patterns, capturing both large-scale trends and finer details.}
    \label{Fig.15A}
\end{figure}

\begin{figure}[htbp]
    \centering
    \includegraphics[width=1\textwidth]{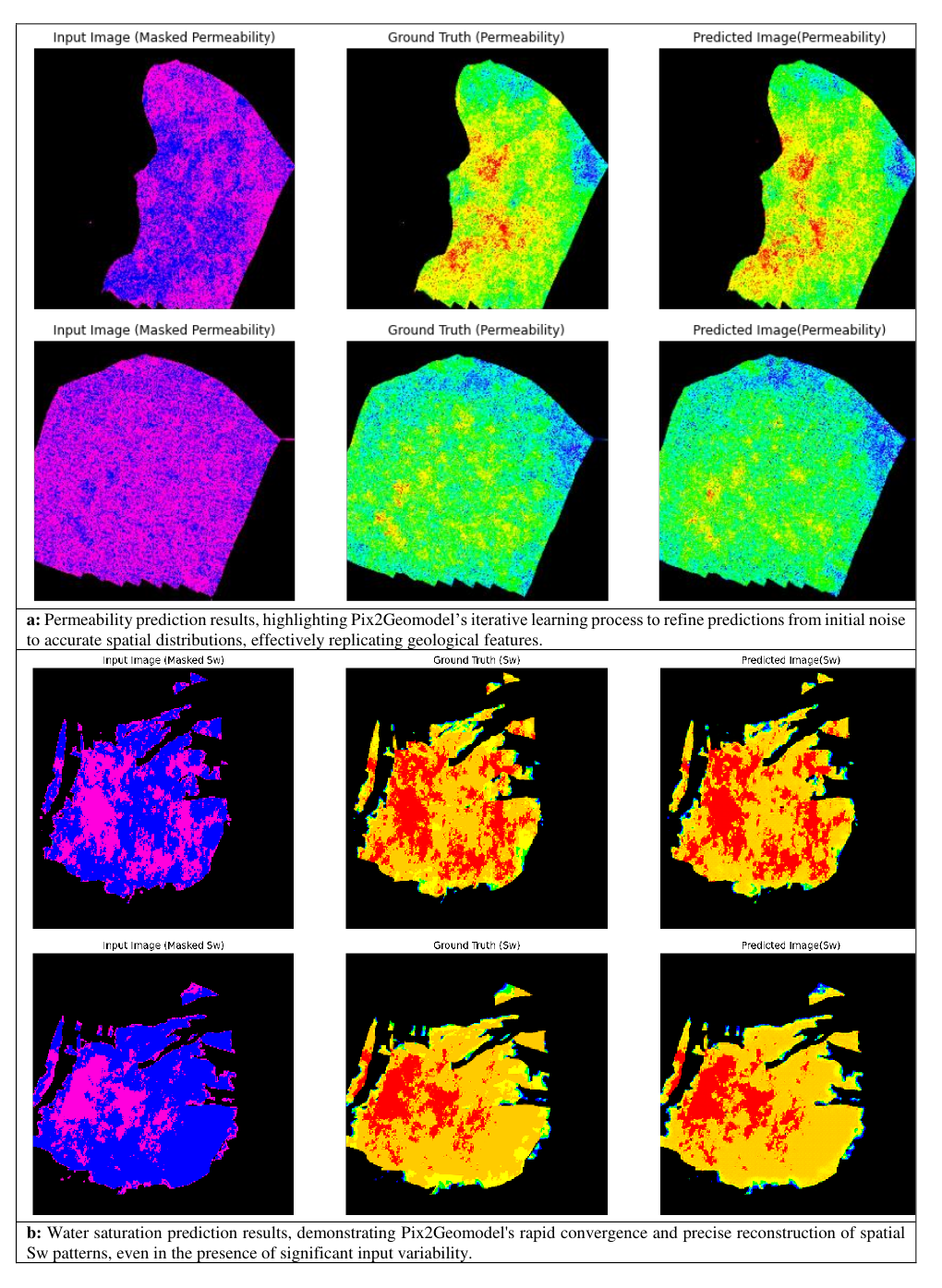}  
    \caption{Pix2Geomodel predictions for permeability (a) and water saturation (b) compared to ground truth. The model demonstrates its robustness in predicting spatial distributions, transitioning from noise-like outputs at early stages to accurate reconstructions with prolonged training.}
    \label{Fig.16A}
\end{figure}

\begin{figure}[htbp]
    \centering
    \includegraphics[width=1\textwidth]{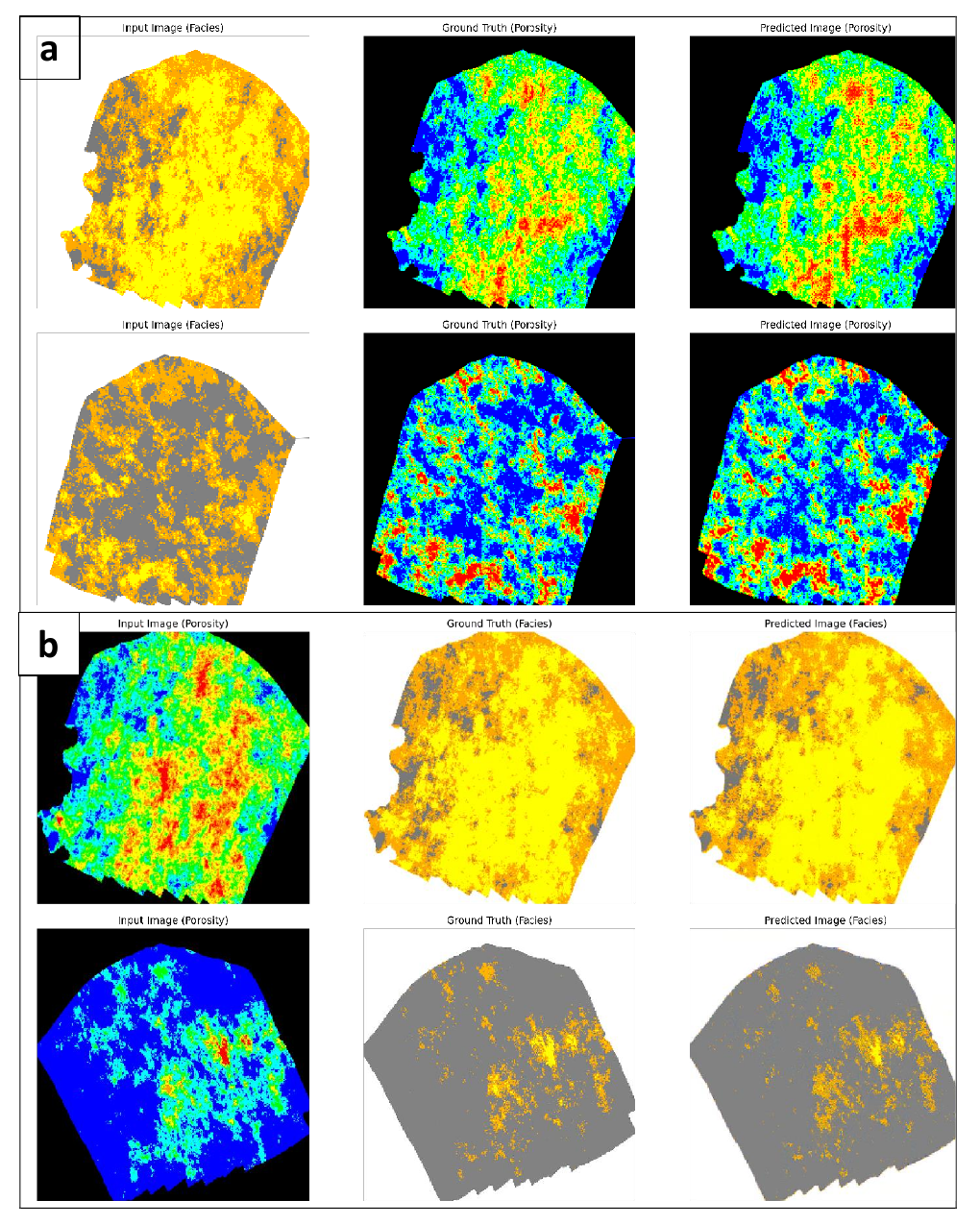}  
    \caption{(a)Facies-to-porosity translation results – Predicted porosity distributions (right) derived from facies input data (left) compared against ground truth porosity (center).(b)Porosity-to-facies translation results – Predicted facies distributions (right) derived from porosity input data (left) compared against ground truth facies (center). }
    \label{Fig.17A}
\end{figure}

\begin{figure}[htbp]
    \centering
    \includegraphics[width=1\textwidth]{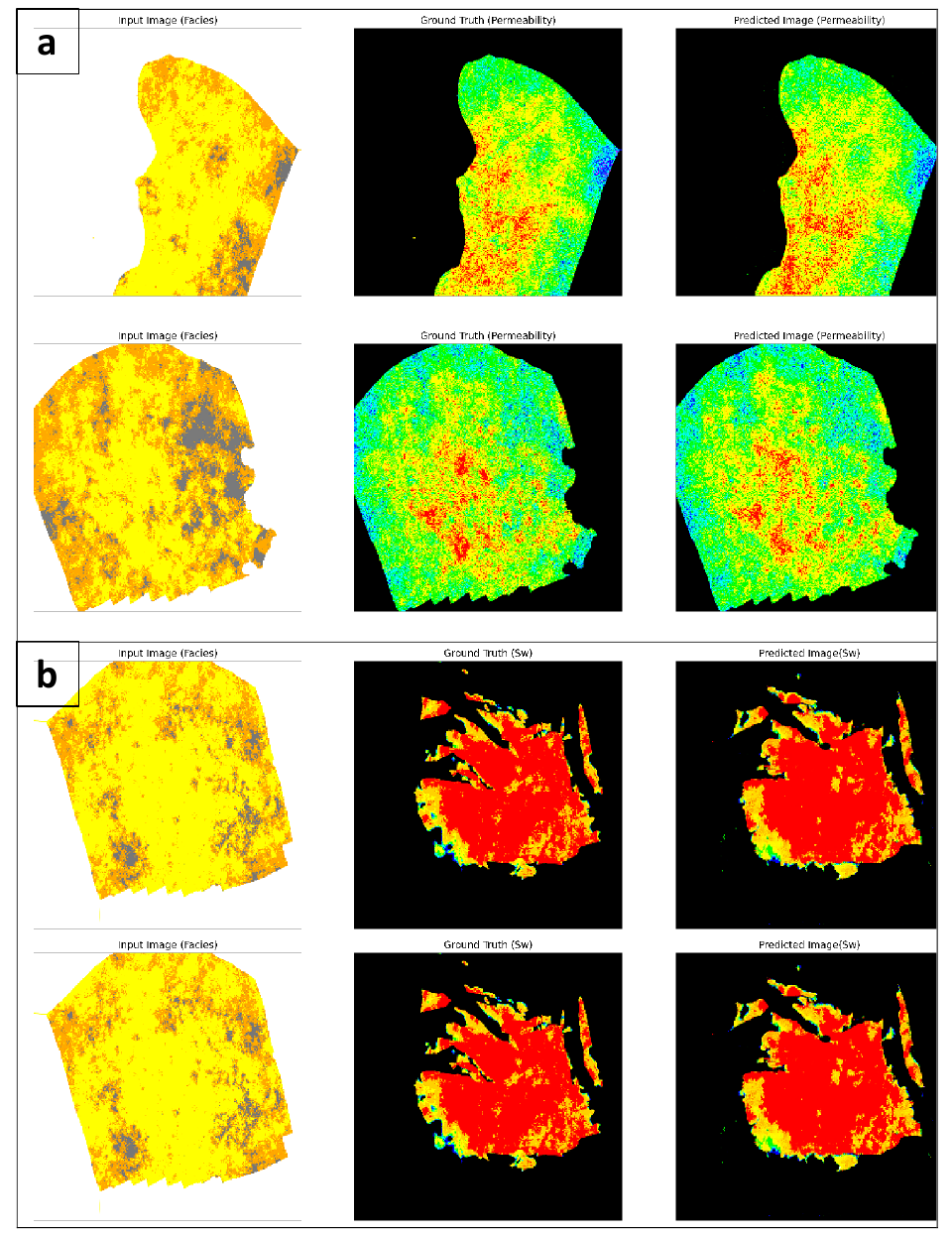}  
    \caption{(a) Facies-to-permeability translation results – Predicted permeability distributions (right) derived from facies input data (left) compared against ground truth permeability (center). (b)Facies-to-Sw translation results – Predicted permeability distributions (right) derived from facies input data (left) compared against ground truth permeability (center).}
    \label{Fig.18A}
\end{figure}

\end{document}